\documentclass[aps, prd, reprint, superscriptaddress, amsmath, amssymb]{revtex4-2}

\usepackage{graphicx}
\usepackage{dcolumn}
\usepackage{bm}
\usepackage{aas_macros}
\usepackage{mathtools}
\usepackage{lipsum} 
\usepackage{xcolor}
\usepackage{float}
\usepackage[normalem]{ulem}

 \usepackage{hyperref}
 \hypersetup{
     colorlinks=true,
     linkcolor=black,
     filecolor=blue,
     citecolor = blue,      
     urlcolor=magenta}

\usepackage[sort&compress]{natbib}
\newcommand{\CosmoDC}{\texttt{CosmoDC2 }}

\begin{document}
\title{Perturbation theory  models for LSST-era galaxy clustering: tests with sub-percent mock catalog measurements in Fourier and configuration space}

\author{Samuel Goldstein}
\author{Shivam Pandey}
\affiliation{Department  of  Physics  and  Astronomy,  University  of  Pennsylvania,  Philadelphia,  PA  19104,  USA}

\author{An\v{z}e Slosar}
\affiliation{Brookhaven National Laboratory, Upton NY 11973}

\author{Jonathan Blazek}
\affiliation{Department  of  Physics,  Northeastern  University,  Boston,  MA  02115,  USA}
\affiliation{Laboratory  of  Astrophysics, \'Ecole  Polytechnique  F\'ed\'erale  de  Lausanne  (EPFL),Observatoire  de  Sauverny,  1290  Versoix,  Switzerland}

\author{Bhuvnesh Jain}
\affiliation{Department  of  Physics  and  Astronomy,  University  of  Pennsylvania,  Philadelphia,  PA  19104,  USA}

\author{the LSST Dark Energy Science Collaboration}

\begin{abstract}
  We analyze the clustering of galaxies using the $z=1.006$ snapshot of the \CosmoDC simulation, a high-fidelity synthetic galaxy catalog designed to validate analysis methods for the Vera C. Rubin Observatory Legacy Survey of Space and Time (LSST). We present sub-percent measurements of the galaxy auto-correlation and galaxy-dark matter cross correlations in Fourier space and configuration space for a magnitude-limited galaxy sample. At these levels of precision, the statistical errors of the measurement are comparable to the systematic effects present in the simulation and measurement procedure; nevertheless, using a hybrid-PT model, we are able to model non-linear galaxy bias with 0.5\% precision up to scales of $k_{\rm max}= 0.5 \ h$/Mpc and $r_{\rm min}= 4$ Mpc/$h$.  While the linear bias parameter is measured with 0.01\% precision, other bias parameters are determined with considerably weaker constraints and sometimes bimodal posterior distributions. We compare our fiducial model with lower dimensional models where the higher-order bias parameters are fixed at their co-evolution values and find that leaving these parameters free provides significant improvements in our ability to model small scale information. We also compare bias parameters for galaxy samples defined using different magnitude bands and find  agreement between samples containing equal numbers of galaxies. Finally, we compare bias parameters between Fourier space and configuration space and find moderate to significant tension between the two approaches. Although our model is often unable to fit the \CosmoDC galaxy samples within the 0.1\% precision of our measurements, our results suggest that the hybrid-PT model used in this analysis is capable of modeling non-linear galaxy bias within the percent level precision needed for upcoming galaxy surveys.
\end{abstract}

\maketitle

\section{Introduction}

Galaxy clustering is becoming an increasingly important tool in the field of observational cosmology. Throughout the next decade, numerous wide-field galaxy surveys such as the Legacy Survey of Space and Time (LSST) from the Vera Rubin Observatory \cite{0912.0201}, Euclid \cite{1001.0061}, the Nancy Grace Roman Space Telescope \cite{1305.5425}, the Dark Energy Spectroscopic Instrument (DESI) \cite{DESI_Part_1}, 4-metre Multi-Object Spectroscopic Telescope (4MOST) \cite{4MOST-Cosmology}, Hawaii Two-0 Survey (H20) \cite{Hawaii_two_0_cosmology}, and the Subaru Prime Focus Spectrograph (PFS) \cite{subaru_pfs} will map the distribution of galaxies with an unprecedented level of precision, complementing existing datasets from the Dark Energy Survey (DES) \cite{1708.01530, descollaboration2021dark}, the Kilo-Degree Survey(KiDS) \cite{KiDS_Cosmic_Shear}, the Hyper-Suprime Cam (HSC) \cite{1809.09148}, and the Baryon Oscillation Spectroscopic Survey (BOSS and eBOSS) \cite{Alam_2017, Alam2021}.  In order to extract the wealth of cosmological information present in galaxy clustering data, these surveys will require a range of advanced modeling techniques. In preparation for this, the cosmology community has made great advances in its ability to simulate synthetic galaxy catalogs \cite{1312.1707,1712.05768,1804.05865,1904.11970,2021ApJS..252...19H,2004.06245} and model summary statistics beyond the linear regime \cite{1910.07097,Takahashi:2012em,pandey2021dark, Kokron_2021}. 

In this paper, we validate the perturbative bias model used in the DES Y3 \cite{2008.05991,descollaboration2021dark,pandey2021dark,porredon2021dark} analysis (both for the sample of red galaxies and the sample of bright magnitude limited galaxies) against a deep, magnitude limited LSST-like galaxy sample from the \CosmoDC simulation, the largest synthetic photometric dataset to date. The less-biased galaxies included in our faint samples are likely to be easier to model than the bright galaxies residing in the centers of the most massive halos, which tend to come from overdense regions of the Universe subject to stronger non-linear gravitational processes. Nevertheless, the increased statistical precision and non-trivial ways in which fainter galaxies occupy halos pose novel challenges to non-linear modeling. 

In addition to testing the DES Y3 bias model on an LSST-like galaxy sample, we compare bias parameters for varying luminosity cuts to study whether the choice of the bands in which we perform the cuts matters. We also analyze two-point statistics in both Fourier and configuration spaces using the same galaxy sample and bias model in order to assess the level of agreement between the two methods. Although large separations in configuration space encode similar information as small wavenumbers in Fourier space, the mapping is only approximate in the presence of sharp scale cuts. Hence results from the two methods are expected to be only partially correlated \cite{Doux_cosmic_shear_real_Fourier} (c.f. ``consensus results" in BAO fitting of spectroscopic data \cite{2018MNRAS.473.4773A}).

This paper is organized as follows. In \S\ref{sec:Data_Methods} we describe the \CosmoDC simulation, our measurements, bias models, and fitting procedure. We present our results in  \S\ref{sec:results} and conclusions in \S\ref{sec:conclusions}.

\section{Data \& methods}\label{sec:Data_Methods}
\subsection{CosmoDC2 simulations and sample selection}

\CosmoDC is a large synthetic catalog of galaxies generated for the LSST Dark Energy Science Collaboration (DESC) to satisfy its simulation needs \cite{1907.06530}. It was created to support the generation of the DC2 dataset \cite{2010.05926, 2101.04855}, but is itself a standalone and public data product. The catalog is based on the \textit{Outer Rim} \cite{1904.11970} simulation, a pure dark matter $N$-body simulation consisting of $10,240^3$ particles in a $27 \ ({\rm Gpc}/h)^3$ box with mass resolution $\sim 2.6 \times 10^{9}M_\odot$ and cosmological parameters close to the best-fit WMAP-7 set \cite{komatsu_2011} with $\omega_{\rm cdm}=0.1109,$ $\omega_{\rm b}=0.02258,$ $n_s=0.963,$ $h=0.71,$ $\sigma_8=0.8,$ and $w=-1.0$. \CosmoDC covers $440 \ \rm deg^2$ of sky area to a redshift of $z=3$ and is complete to a magnitude depth of 28 in the $r$-band. Galaxies were populated based on a hybrid technique combining empirical models using Universe Machine \cite{1806.07893} with semi-analytic models based on the Galacticus \cite{1008.1786} code. \CosmoDC was calibrated against observed properties using the DESCQA validation tool \cite{1709.09665} and the final product is a galaxy catalog that is realistic in a number of galaxy properties, including stellar mass, morphology, spectral energy distributions, broad-band filter magnitudes, host halo information, and weak lensing shear. It is currently not feasible to run hydrodynamic codes at these volumes and precision, but within the limitations of the HOD \cite{BerlindWeinberg02, Zheng05} framework, \CosmoDC represents one of the most realistic large-volume galaxy catalogs available. In this analysis we work directly with the galaxy distribution in real-space and leave the observational complications (curved sky, selection functions, blending, etc.) for future work.

We analyze \CosmoDC snapshot data at redshift $z=1.006$, which is near the peak of large scale structure constraining power for the LSST survey. Our primary galaxy sample consists of three magnitude limited samples corresponding to $r$-band thresholds of 22, 23, and 24.5. We also analyze galaxies with $i$-band limiting magnitudes of 21.80, 22.87, and 24.43 corresponding to an equal number of galaxies as the $r$-band cuts.

\begin{figure}[t]
\includegraphics[width=\linewidth]{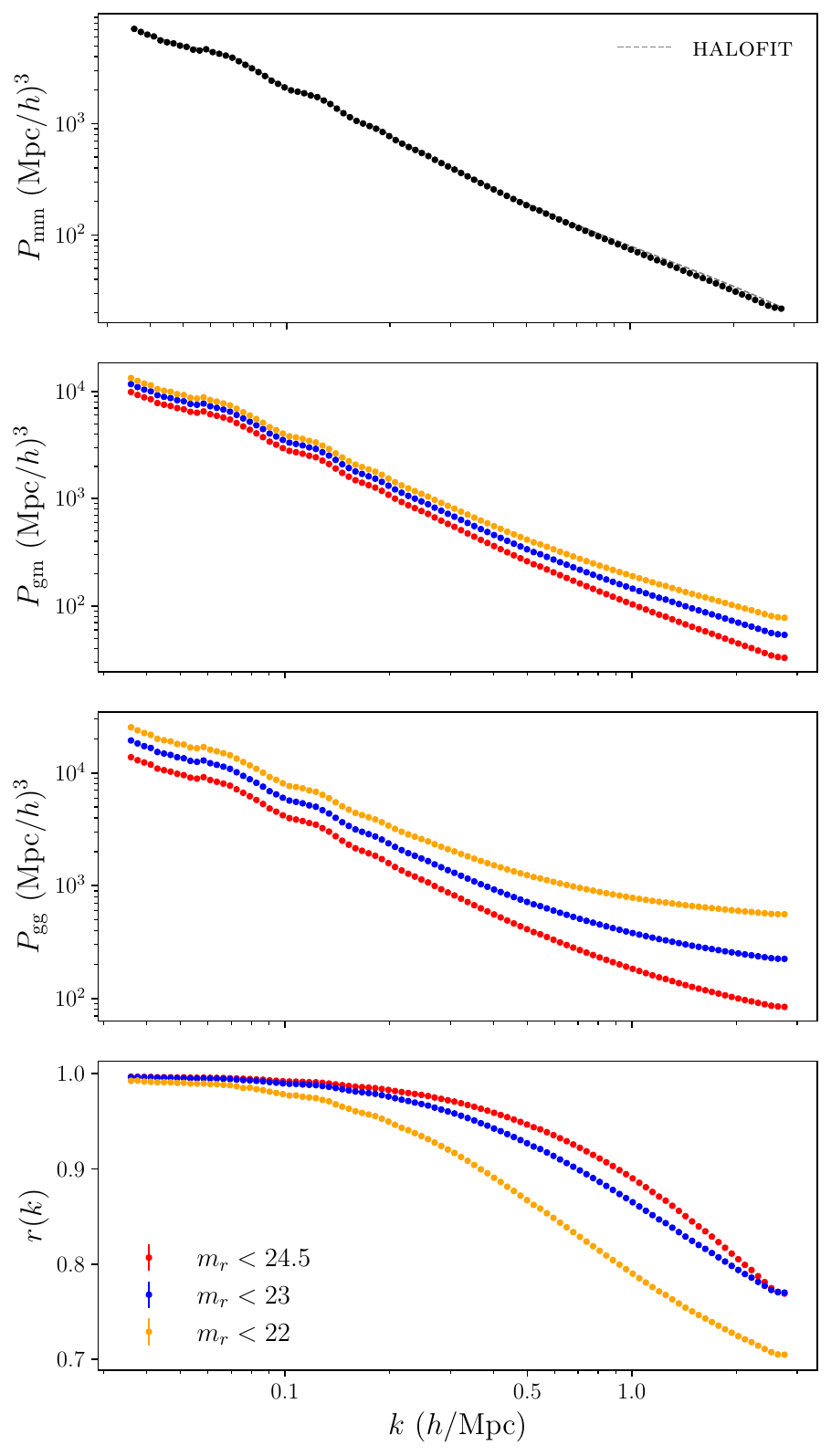}
\caption{Measurements of the 3D matter ($P_{\rm mm}$), galaxy-matter ($P_{\rm gm}$), and galaxy ($P_{\rm gg}$) power spectra for \CosmoDC galaxies at redshift $z=1.006$. The gray-dashed curve in the top panel shows the \textsc{halofit} \citep{Takahashi:2012em} non-linear matter power spectrum estimate. The galaxy-matter cross correlation coefficients ($r$) are shown in the bottom panel. Shot-noise is not subtracted from $P_{\rm gg}$.}
\label{fig:figure1}
\end{figure}

\subsection{Power spectrum measurement}\label{sec:measure_Pk}
We compute the galaxy power spectrum ($P_{\rm gg}$), the galaxy-matter cross power spectrum ($P_{\rm gm}$), and the matter power spectrum ($P_{\rm mm}$) using the publicly available \texttt{nbodykit} package \cite{Hand_2018}. We interpolate the galaxy and dark matter density fields of the entire simulation volume onto a $1536^3$ mesh using a Cloud In Cell assignment scheme and compute the relevant power spectra in 300 logarithmically spaced bins between $0.01$ and $2.8 \ h/{\rm Mpc}$. We then remove two Fourier modes with excess power as described in Appendix \ref{Appendix:Excess_Pow} and rebin the measured power spectra into 80 logarithimically spaced bins between $0.035$ and $2.8 \ h/{\rm Mpc}$.

We estimate the power spectra covariance using the subsample covariance obtained by dividing the entire simulation volume into 64 equally sized subboxes and computing $P_{\rm gg}, \ P_{\rm gm},$ and $P_{\rm mm}$ within each subbox. The covariance matrix is then rescaled to apply to the full simulation volume of $27 \ ({\rm Gpc}/h)^3$. We inflate the diagonal of our covariance estimate by 5\% to ensure stability of our model fits which require inverting the covariance matrix. Note that our covariance does not account for the non-periodicity of the subboxes when computing the power spectra; however, we expect the impact of this to be small on the scales of our analysis ($0.035$ to $0.9 \ h/{\rm Mpc}$) as the subboxes have a boxsize of 750 ${\rm Mpc}/h$. Moreover, these non-periodic effects should largely cancel out in our final analysis in which we consider only the ratios $P_{\rm gg}/P_{\rm mm}$ and $P_{\rm gm}/P_{\rm mm}$. A detailed discussion of our covariance matrix estimation techniques and their implications on our results is included in Appendix \ref{Appendix:Covariance_Adjustment}. 

In Figure \ref{fig:figure1} we show the computed power spectra for the three $r$-band limited galaxy samples, as well as the galaxy-matter cross correlation coefficient $r \equiv {P_{\rm gm}}/{\sqrt{P_{\rm mm}P_{\rm gg}}}$. We do not subtract shot-noise from $P_{\rm gg}$ in this figure.

\subsection{Correlation function measurement}\label{sec:measure_xi}

In addition to measuring the power spectrum, we measure and analyze the two-point correlations between galaxies and matter particles in configuration space. In order to reduce the computational memory required for these correlation function measurements, we implement a voxel based correlation function estimator. This method removes the computational complexity of estimating two-point correlations of a large random catalog (approximately equal to $10$ times the total number of galaxies, $N_{\rm total}$) in the typical Landy-Szalay estimator \cite{Landy_Szalay93}. We divide our simulation box (with volume $V_{\rm total}$) into 3D cubes (called voxels) of volume $V_{\rm voxel} = 1 \ ({\rm Mpc}/h)^3$. Thereafter, we assign weights to each voxel $i$, given by $w_i = (N_i/V_{\rm voxel})\times (V_{\rm total}/N_{\rm total})$ where $N_i$ is the number of galaxies inside voxel $i$. Our voxel based estimator can then be written as:
\begin{equation}
    \xi(r) = \sum_i \sum_j (w_i - 1)(w_j - 1) \Theta_{i,j}
\end{equation}
where $\Theta_{i,j}$ is a top-hat like function that is equal to unity when the voxels $i$ and $j$ are separated by 3D distance $r$, within the bin-width of $\Delta r$ (calculated from the binning). We measure the correlation function in 20 bins ranging from $1.9 \ {\rm Mpc}/h$. to $60 \ {\rm Mpc}/h$. We have verified that this voxel based estimator and regular Landy-Szalay based estimator using randoms are negligibly different within our error bars.  We note that the configuration space measurements are obtained using only half of the total simulation volume because of computational constraints.

We estimate the configuration space covariance using the jackknife method \cite{QUENOUILL_56, Tukey_58} by dividing the simulation into 512 regions. We only use scales below $40 \ {\rm Mpc}/h$ to fit our models, where we have a reliable error estimate. We rescale the covariance to apply to the full simulation volume of $27 \ ({\rm Gpc}/h)^3$. In order to improve the stability of the inverse covariance (when calculating the likelihood), we inflate the diagonal elements of the covariance matrix by 5\%. We have verified that voxel size does not affect the correlation function measurement beyond 4 Mpc$/h$ by repeating the measurement with a voxel twice the fiducial size.

\subsection{Perturbative bias model}\label{sec:PT_Models}

The perturbation theory framework aims to describe the overdensity of a biased tracer of dark matter, such as galaxies, in terms of the matter overdensity. This relationship is encoded in the bias parameters. In this analysis we typically work on scales larger than the Lagrangian radius of the host halos of our galaxies (denoted by $R_{*}$) which is the radius in early Universe Lagrangian space from which the matter accretes inside the halo. On account of this, alongside the fact that the large-scale growth factor is scale independent, we work under the approximation that the galaxy overdensity, $\delta_{\rm g}$, can be described as a function of matter density at the same redshift (see \citet{Desjacques_2018} for a detailed review). 

At large scales and high redshift, the physics of overdensity perturbations is largely linear, and hence galaxy bias is well approximated by a linear relation. The gravitational evolution of the dark matter naturally results in non-linear and non-local effects which become dominant at smaller scales and lower redshifts. Assuming homogeneity and isotropy, it can be shown that these non-linear and non-local terms can only be sourced by scalar quantities constructed out of gravitational evolution of matter density ($\delta_{\rm m}$), shear ($\nabla_i \nabla_j \phi$, where $\phi$ is gravitation potential), and velocity divergences ($\nabla_i v_j$, where $v_j$ is the $j$-th component of the 3D particle velocity). As described in \citep{McDonald2009, Chan_2012}, the expansion of the galaxy overdensity ($\delta_g$) can be re-arranged into independent terms that contribute at different orders:

\begin{multline}\label{eq:delg_full}
    \delta_{\mathrm{g}} \sim f(\delta_{\mathrm{m}}, \nabla_i \nabla_j \Phi, \nabla_i v_j) \sim f^{(1)}(\delta_{\mathrm{m}}) + f^{(2)}(\delta^2_{\mathrm{m}}, s^2)  \\ + f^{(3)}(\delta^3_{\mathrm{m}},\delta_{\mathrm{m}} s^2, \psi, st) + ....
\end{multline}

\noindent where, $f^{i}$ are functions that contribute to the total overdensity at $i$-th order only and $\psi,s$ and $t$ are scalar quantities constructed out of shear and velocity divergences. Note that these terms are all spatially local, meaning that galaxy overdensity at any Eulerian position is expressed in terms of the matter density evaluated at same position. Nevertheless, galaxy formation is a non-local process in which matter from nearby areas collapses. As described in \citet{McDonald2009}, the lowest order contribution from this process is captured by the Laplacian of the matter overdensity, $\nabla^2 \delta_{\rm m}$. We incorporate this term in our theory model as well.

An explicit expansion of Eq.~\ref{eq:delg_full} in terms of matter over-density $\delta_{\rm m}$ introduces a set of ``bare-bias" parameters that are unobservable and can not necessarily be attributed a physical interpretation. At the power spectrum level, a re-normalization of these ``bare-bias" parameters can be performed by combining terms with similar kernels (see \citet{McDonald2009} for a detailed calculation). After re-normalizing, we can write the tracer-matter cross spectrum ($P_{\mathrm{gm}}$) and the tracer auto power spectrum ($P_{\mathrm{gg}}$) as:

\begin{multline}\label{eq:P_tm}
P_{\mathrm{gm}}(k)= b_1 P_{\mathrm{mm}}(k) +  \frac{1}{2} b_2P_{\rm b_1 b_2}(k) + \frac{1}{2} b_{\mathrm{s}}P_{\rm b_1 s^2}(k) + \\ \frac{1}{2} b_{\rm 3nl}P_{\rm b_1 b_{\rm 3nl} }(k) + b_{k}  k^2 P_{\mathrm{mm}}(k) 
\end{multline}

\begin{multline}\label{eq:P_tt}
	P_{\mathrm{gg}}(k) = b_1^2 P_{\mathrm{mm}}(k) + b_1b_2 P_{\rm b_1 b_2}(k) + b_1b_{\mathrm{s}}P_{\rm b_1 s^2}(k) + \\ b_1b_{\rm 3nl}P_{\rm b_1 b_{\rm 3nl} }(k) +  \frac{1}{4}b_2^2 P_{\rm b_2 b_2}(k) + 
	\frac{1}{2}b_2b_{\mathrm{s}}P_{\rm b_2 s^2}(k) + \\ \frac{1}{4}b_{\mathrm{s}}^2 P_{\rm s^2 s^2}(k)  + 2 b_1  b_{k} k^2 P_{\mathrm{mm}}(k) 
\end{multline}

\noindent where $b_1, b_2, b_{\rm s}, b_{\rm 3nl}$ and $b_{\rm k}$ are the re-normalized bias parameters.

This five parameter 1-loop perturbation theory model is complete up to third order in its dependence on the matter overdensity and includes the higher-order bias contribution arising from non-local galaxy formation. The power spectrum $P_{\rm b_1 b_2}(k)$ is generated from the ensemble average of $\langle \delta_{\mathrm{m}} \delta^2_{\mathrm{m}} \rangle$, $P_{\rm b_1 s^2}(k)$ is generated from $\langle \delta_{\mathrm{m}} s^2 \rangle$ and the kernel $P_{\rm b_1 b_{\rm 3nl} }$ is generated from a combination of the ensemble average between $\delta_{\mathrm{m}}$ and arguments of $f^{(3)}$ (see Eq.~\ref{eq:delg_full}) that contribute at 1-loop level \citep{Saito2014a}. These terms involve convolution of the linear matter power spectrum with various kernels and we refer the reader to Appendix A of \citet{Saito2014a} for the form of these kernels. The sum of the higher-order bias terms that are not directly coupled to $P_{\rm mm}(k)$ gives the 1-loop corrections $P_{\rm gg}^{\rm 1-loop}(k)$ and $P_{\rm gm}^{\rm 1-loop}(k)$. The scale-dependent term including $k^2P_{\rm mm}(k)$ originates from higher derivative bias. 

This model does not include the 2-loop corrections that are expected to become significant at scales around $k\approx0.3$ $h$/Mpc for $z\approx 1$ \citep{Senatore_IR}; however, by using a hybrid PT model in which we model $P_{\rm mm}$ using \textsc{halofit} simulation based fitting formula detailed in \citep{Takahashi:2012em}, we expect to partially capture the 2-loop contributions. Furthermore, this non-linear matter power spectrum estimate should capture the counter-terms described by the effective field theory \citep{Carrasco:2012cv, Vlah15, perko2016biased}. We defer the analysis incorporating all 2-loop terms in galaxy biasing to a future study. In Appendix \ref{Appendix:Halo_Fit} we compare our measured matter power spectrum with the \textsc{halofit} estimate and find they agree within 5\% which is sufficient for this analysis.

An alternative approach to describe the clustering of biased tracers is to use the Lagrangian formulation (see \citet{Matsubara_2013}), which, assuming zero contribution from higher derivative biases, results in an equivalent description of $P_{\rm gg}$ and $P_{\rm gm}$ at 1-loop order. This allows us to infer the values of the bias parameters $b_{\rm s}$ and $b_{\rm 3nl}$. These co-evolution values depend only on the linear bias parameter $b_1$ and Lagrangian space contributions and are given by $b_{\rm s}  \equiv -(4/7)\times(b_1-1) + b^{\rm Lag}_{\rm s}$ and  $b_{\rm 3nl}\equiv (b_1-1) + b^{\rm Lag}_{\rm 3nl}$. Here $b^{\rm Lag}_{\rm s}$ and $b^{\rm Lag}_{\rm 3nl}$ are the bias values at high redshift of Lagrangian space. In this analysis we assume $b^{\rm Lag}_{\rm s} = b^{\rm Lag}_{\rm 3nl} = 0$, but test the deviation from this assumption when analyzing the data as described in the following section (also see \cite{Modi_2017}).

\subsection{Models}

We consider three models of decreasing complexity based on the bias models in Eq.~\ref{eq:P_tm} and Eq.~\ref{eq:P_tt}:
\begin{itemize}
    \item \textit{Fiducial model:} $b_1, b_2, b_s, b_{\rm 3nl},$ and $b_{\rm k}$ vary freely
    \item \textit{3-parameter model:} $b_1, b_2,$ and $b_{\rm k}$ vary freely with $b_s$ and $b_{\rm 3nl}$ fixed at their co-evolution values
    \item \textit{2-parameter model:} $b_1$ and $b_2$ freely with $b_s$ and $b_{\rm 3nl}$ at their co-evolution values and $b_{\rm k}$ set to zero
\end{itemize}

We use the publicly available FAST-PT code \citep{McEwen_2016, Fang_2016} as implemented in the Core Cosmology Library (CCL) \citep{Chisari_2018} to evaluate the mode coupling integrals in Eq.~\ref{eq:P_tm} and Eq.~\ref{eq:P_tt}. For predictions of the matter power spectrum, we use the updated version of \textsc{halofit} \citep{2003MNRAS.341.1311S, Takahashi:2012em}.

\subsection{Fitting procedure}\label{subsec:fitting_procedure}

\begin{figure*}[!htbp]
\centering
\includegraphics[width=\linewidth]{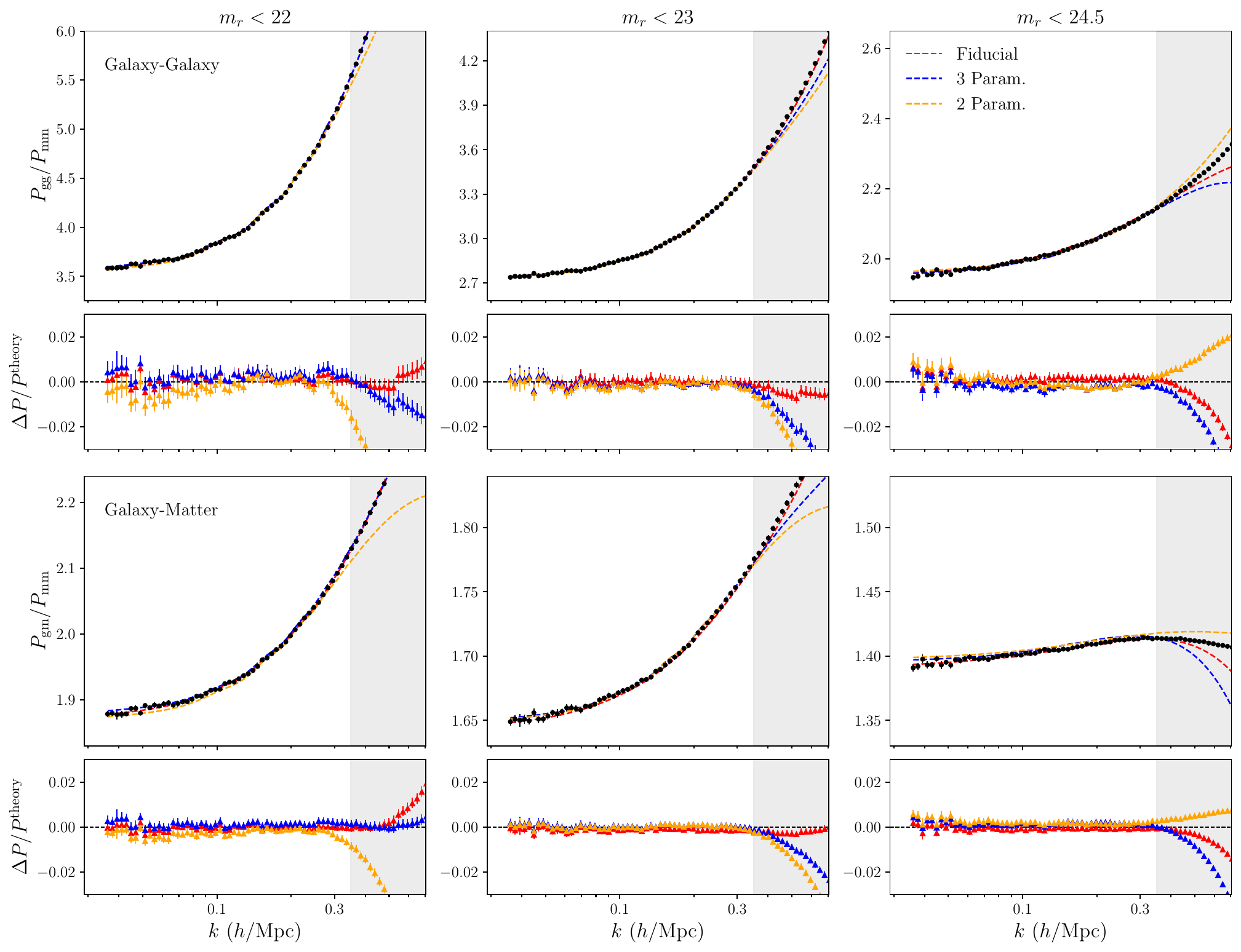}
\caption{ $P_{\rm gg}/P_{\rm mm}$ and $P_{\rm gm}/P_{\rm mm}$ joint fit results for the three $r$-band galaxy samples with $k_{\rm max}= 0.35 \ h/\rm{Mpc}$ (gray region). The residuals ($\Delta P/P^{\rm theory}$) are shown in the panels below each model fit.}

\label{fig:ratio_fits}
\end{figure*}

\subsubsection{Power spectrum}
In a traditional analysis, one would fit the matter-matter, galaxy-matter and galaxy-galaxy power spectra jointly. Although translational invariance in the periodic box ensures that power spectra at different wavenumbers are independent at large scales, the three power spectra at the same $k$ are not. In particular, since the three power spectra are physically tracing the same structure, measurements of $P_{\rm mm}, P_{\rm gm},$ and $P_{\rm gg}$ are heavily correlated, leading to a poorly conditioned $3\times3$ covariance matrix. Furthermore, at the extreme precision of the simulation at hand, the \textsc{halofit} model itself is a limited approximation of the matter power spectrum. It has also been found that galaxy power spectra of mocks generated from the \textit{Outer Rim} simulation in \citet{2106.12580} are not well fit by the fiducial cosmology even for relatively conservative scale cuts. Since we are interested in the model of galaxy tracing we avoid these problems by working with the ratios $P_{\rm gm}/P_{\rm mm}$ and $P_{\rm gg}/P_{\rm mm}$. These quantities are still correlated at the same wavenumber, but with a considerably more diagonal matrix. Additionally, when analyzing ratios, the shape of the non-linear power spectrum enters only through the loop corrections of higher order terms. The theory model for these ratios is given by

\begin{equation}\label{eq:gmomm} 
\frac{P_{\rm gm}}{ P_{\rm mm}} = b_1+b_{\rm k} k^2 + \frac{P_{\rm gm}^{\rm 1-loop}(k)}{P_{\rm mm}^{HF}(k)}\\
\end{equation}

\begin{equation}\label{eq:ggomm} 
\frac{P_{\rm gg}}{P_{\rm mm}} = b_1^2 + 2b_1 b_{\rm k} k^2 + \frac{P_{\rm gg}^{\rm 1-loop}(k)}{P_{\rm mm}^{HF}(k)}+\frac{1+N_0}{\overline{n}_{\rm g} P_{\rm mm}^{HF}(k)}\\  
\end{equation}

 \noindent where $P_{\rm mm}^{HF}(k)$ is the \textsc{halofit} non-linear matter power spectrum. The form of $P_{\rm gg}^{\rm 1-loop}(k)$ and $P_{\rm gm}^{\rm 1-loop}(k)$ can be obtained by comparing the numerators of the above equations to Eq.~\ref{eq:P_tm} and Eq.~\ref{eq:P_tt}. The final term of Equation \ref{eq:ggomm} allows for scale-independent deviations from Poissonian shot-noise, given by inverse number density ($1/\overline{n}_{\rm g}$), via the free parameter $N_0$. In Appendix \ref{Appendix:poisson_shotnoise} we include results assuming Poissonian shot-noise ($N_0=0$).

 We fit the ratios $P_{\rm gm}/P_{\rm mm}$ and $P_{\rm gg}/P_{\rm mm}$ jointly assuming a Gaussian likelihood using the covariance described in \S\ref{sec:measure_Pk}. We adopt uniform priors ranging from 0 to 2 for $b_1$, $-10$ to $10$ for the non-linear bias parameters, and $-1$ to $1$ for the non-Poissonian shot-noise, $N_0$. We obtain samples from the bias parameter posterior distributions using the publicly available \texttt{MultiNest} package \cite{Multinest}. In order to evaluate the dependence of bias parameters on the choice of scale cuts, we run this fitting procedure for six magnitude limited galaxy samples at sixteen scale cuts linearly spaced between $0.15$ and $0.9 \ h/\rm{Mpc}$.
 
\begin{figure*}
\centering
\includegraphics[width=\linewidth]{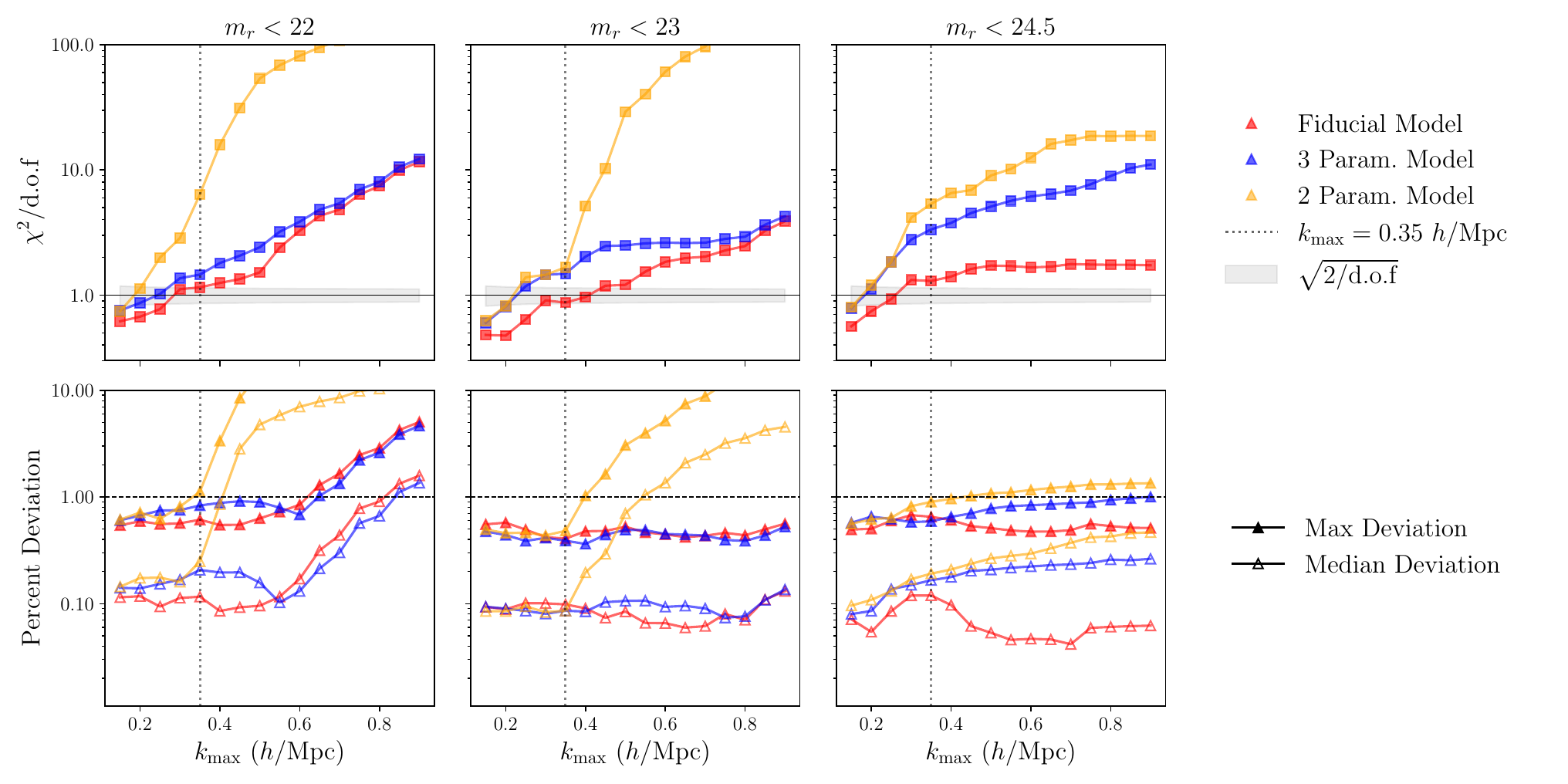}
\caption{\textit{Top}: Reduced $\chi^2$ as a function of $k_{\rm max}$ for the models described in \S\ref{sec:PT_Models} for all $r$-band galaxy samples. The gray band brackets the expected scatter in the reduced $\chi^2$ for the fiducial model. We omit the regions for the two and three parameter models as they deviate by less than 5\% from that of the fiducial model. \textit{Bottom}: Maximum (filled) and median (unfilled) percentage deviation as a function of $k_{\rm max}$ for each bias model. The 1\% region is indicated using a horizontal black dashed line.}
\label{fig:Fourier_space_reduced_chi2}
\end{figure*}
\begingroup
\setlength{\tabcolsep}{5pt} 
\renewcommand{\arraystretch}{1.2}
\begin{table*}[htbp]
    \centering
     \begin{tabular}{|c|| c | c|  c| c| c| c | c| c| c|} 
     \hline
     $m_r$ & $b_1$  & $b_2$ & $b_s$ & $b_{\rm 3nl}$ & $b_{\rm k}$ & $N_0$ & $\chi^2/{\rm d.o.f}$ &$\rm{Max} \ \%$ & $\rm{Median} \ \%$\\ [0.5ex] 
    \hline
      24.5 & 1.3911$^{+0.0005}_{-0.0005}$ & 0.034$^{+0.016}_{-0.020}$ & 0.904$^{+0.037}_{-0.049}$ & 0.91$^{+0.07}_{-0.06}$ & $-0.18$ $^{+0.04}_{-0.04}$  & 0.04 $^{+0.02}_{-0.02}$ & 1.29 & 0.65 & 0.12 \\
     \hline
     23 & 1.6431$^{+0.0007}_{-0.0009}$ & 0.199$^{+0.024}_{-0.028}$ & 0.54$^{+0.08}_{-0.08}$ &1.32$^{+0.10}_{-0.09}$ & 0.44$^{+0.06}_{-0.06}$ &  $-0.08^{+0.01}_{-0.01}$ & 0.87 & 0.41 & 0.10 \\ 
     \hline
     22 & 1.8702$^{+0.0019}_{-0.0013}$ & 0.397$^{+0.054}_{-0.055}$ & 0.20$^{+0.35}_{-0.42}$ & 1.74$^{+0.18}_{-0.25}$ & 1.30$^{+0.12}_{-0.11}$ &
      $-0.07^{+0.01}_{-0.01}$ & 1.15 & 0.61 & 0.12 \\
     \hline
    \end{tabular}
    \caption{Best fit bias parameters and goodness of fit statistics for the fiducial model fits assuming 
    $k_{\rm max} = \ 0.35 \ h/{\rm Mpc}$.}
    \label{tab:b_param} 
\end{table*}
\endgroup

 \subsubsection{Correlation function}

Similarly, we measure and fit the ratios of configuration space correlations, $\xi_{\rm gg}/\xi_{\rm mm}$ and $\xi_{\rm gm}/\xi_{\rm mm}$. Here the configuration space correlation functions are related to the power spectra as $\xi_{xy}(r) = \mathcal{F}^{-1}(P_{xy}(k))$ for $xy \in [{\rm gg},{\rm gm},{\rm mm}]$, and $\mathcal{F}^{-1}$ is the inverse Fourier transform. Note that the power spectrum $k^2 P_{\rm mm}(k)$ in Eq.~\ref{eq:P_tm} and Eq.~\ref{eq:P_tt}, contributed by the higher-derivative bias term, leads to a divergent inverse Fourier transform. Moreover, the other kernels appearing in Eq.~\ref{eq:P_tm} and \ref{eq:P_tt} obtained using FFT-Log method \citep{McEwen_2016} are unstable at higher $k$ values ($k \geq 100 \ h/{\rm Mpc}$), leading to a spurious high-frequency oscillatory feature in the configuration space theory predictions. Therefore we regularize the PT predictions of $P_{\rm gg}$ and $P_{\rm gm}$ by applying an exponential cutoff, $\exp[-(k/k_{*})^4]$, where we choose $k_* = 10  \ h/{\rm Mpc}$. We verify that our conclusions are insensitive to the choice of $k_*$ in Appendix \ref{Appendix:regularization}. We fit the configuration space ratios for the $m_r<23$ galaxy sample at a selection of scale cuts between 4 and 12 ${\rm{Mpc}}/h.$

\section{Results}\label{sec:results}

\subsection{Power spectrum results}

In Figure \ref{fig:ratio_fits} we show the results from fitting the ratios  $P_{\rm gg}/P_{\rm mm}$ and $P_{\rm gm}/P_{\rm mm}$ jointly for three $r$-band galaxy samples assuming ${k_{\rm max}}=0.35 \ h/{\rm Mpc}$. For each magnitude limit, we plot the ratios $P_{\rm gg}/P_{\rm mm}$ (top) and $P_{\rm gm}/P_{\rm mm}$ (bottom) with the theory predictions and residuals for the three models introduced in \S\ref{subsec:fitting_procedure}. We find that the fiducial model produces noticeably better fits than the two and three parameter models, particularly for the $m_r<22$ galaxy sample. The fiducial model is capable of fitting the Fourier space ratios of all galaxy samples within 0.5\% and the two and three parameter models still yield sub-percent level fits. In Table \ref{tab:b_param} we list the best fit bias parameters, reduced $\chi^2$, and maximum and median percent deviation statistics, as defined in the following section, for the fiducial model assuming $k_{\rm max} = 0.35 \ h/\rm{Mpc}$.

\begin{figure*}
\centering
\includegraphics[width=\linewidth]{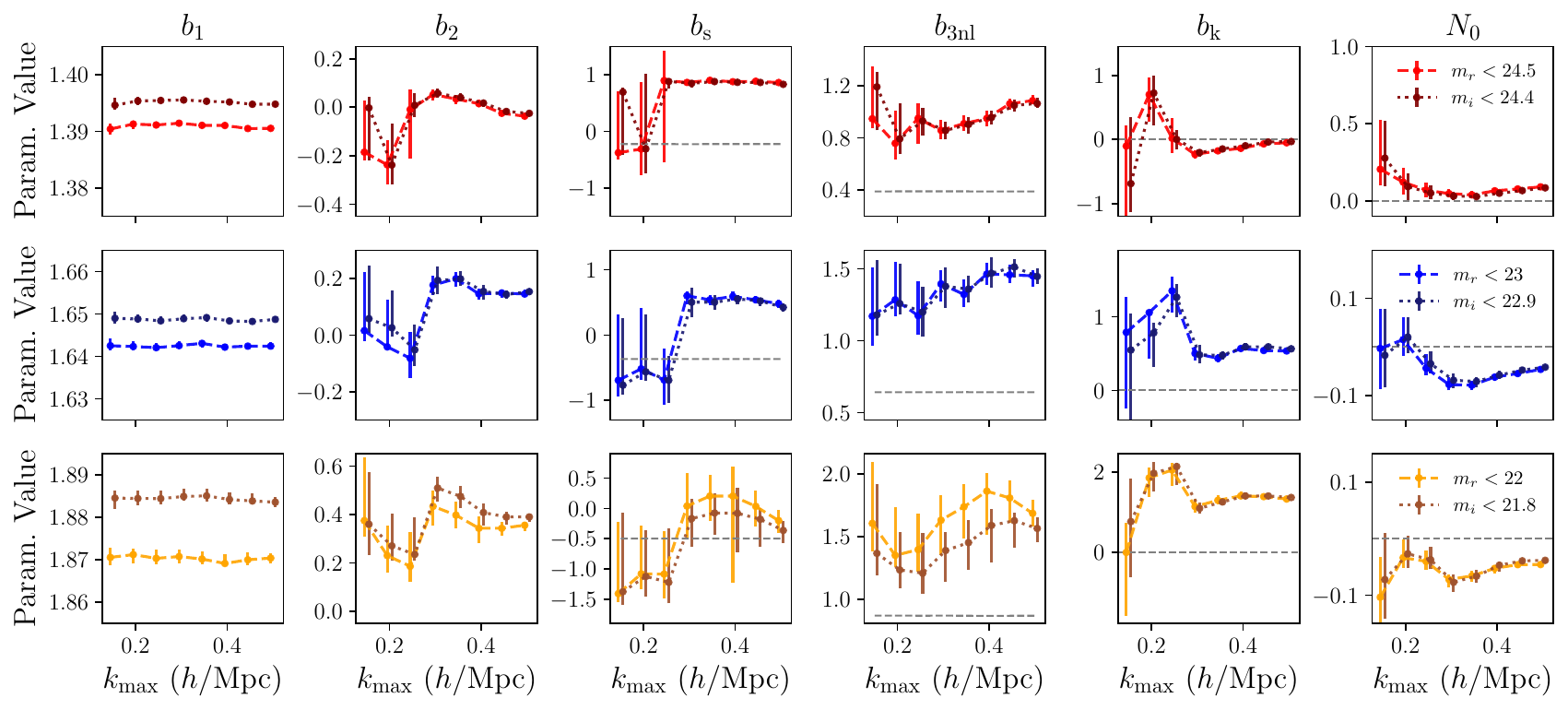}
\caption{Bias parameters as a function of ${k_{\rm max}}$ for our fiducial model fits in Fourier space for all galaxy samples. The dashed lines for $b_s,$ $b_{\rm 3nl}$, and $b_k$ indicate the predictions using the co-evolution relations and $N_0=0$ (Poissonian approximation).}
\label{fig:bias_param_kmax}
\end{figure*}

\subsubsection{Goodness of fit and model comparison}\label{subsec:chi2_comparison}

In order to evaluate our fitting procedure and compare the effectiveness of our bias models across galaxy samples we study the reduced $\chi^2$, maximum percentage deviation, and median percentage deviation as a function of $k_{\rm max}.$ For a given model with $n_p$ free parameters fit using $n_d$ data points, the reduced $\chi^2$ is given by the ratio $\chi^2/\rm{d.o.f}$ where ${\rm d.o.f}=n_d-n_p$ is the total number of degrees of freedom. Whereas the reduced $\chi^2$ statistic is used ubiquitously as an evaluation of goodness of fit, given the high precision of our measurement and artificial inflation of our covariance matrix, perhaps more relevant figures of merit for this analysis are the maximum and median deviation of a model from the data. If a model provides a formally bad $\chi^2$ with respect to the \CosmoDC data, but fits the data with percent level accuracy, then it could still be used in a real analysis provided the measurement error bars are greater than percent level.

Figure \ref{fig:Fourier_space_reduced_chi2} shows the reduced $\chi^2$ and maximum and median percentage deviation as a function of $k_{\rm max}$ for all $r$-band galaxy samples and bias models. Based on the reduced $\chi^2$ values, the fiducial model performs best for the $m_r<24.5$ and $m_r<23$ galaxy samples; however, even for these samples, $\chi^2/\rm{d.o.f}$ is greater than unity for $k_{\rm max}>0.4 \ h/{\rm Mpc},$ indicating that the fit has not fully captured the data given our estimated covariance. The reduced $\chi^2$ values for the fiducial model fits to the $m_r<22$ and $23$ samples are comparable for all scale cuts.

Although our $\chi^2$ values are not a reliable metric for quantifying the effectiveness of a particular model, we can compare $\chi^2$ values across models to assess their relative performance.  In doing so we find that the fiducial model $\chi^2$ values are lower than those of the two and three parameter models for all galaxy samples and scale cuts, hence we can more effectively model small scale information by allowing all bias parameters to vary freely. The difference in performance between the fiducial and three parameter models is less pronounced for the $m_r<22$ and $23$ galaxy samples than the $m_r<24.5$ sample. Furthermore, we find a significant difference in the $\chi^2$ values of the two and three parameter models for the $m_r<22$ and $23$ galaxy samples; however, for the $m_r<24.5$ sample, these models yield comparable $\chi^2$ even at the most extreme scale cuts. This is consistent with the observation that the fiducial model fits for the $m_r<24.5$ sample favor $b_k \sim 0$ across all scale cuts as shown in Figure \ref{fig:bias_param_kmax}.

When evaluated using the max and median percent deviation metrics, the fiducial model performs similarly for the $m_r<23$ and $m_r<24.5$ galaxy samples. For these samples the median and maximum deviations are around $0.1 \%$ and $0.5\%$, respectively, regardless of scale cut. The fiducial model fits for the $m_r<22$ sample are significantly less accurate with median deviation around $0.5\%$ and maximum deviation reaching $4\%$ for $k_{\rm max} \geq 0.4 \ h/{\rm Mpc}.$ The increased difficulty of modeling this sample is expected given the non-linear evolution of the massive halos in which these bright galaxies reside.

Our models generally perform better when applied to the fainter galaxy samples. With the addition of smaller scale information, including $b_{\rm k}$ as a free parameter increases in our model fits for the $m_r<22$ and $m_r<23$ samples, but produces little effect on the $m_r<24.5$ sample suggesting that this galaxy population has less non-local sensitivity. Furthermore, letting all five bias parameters vary freely yields significant improvements in our ability to model all galaxy samples. Our findings differ from those of \cite{2008.05991} who found that the 2-parameter model provides sufficient precision for DES analysis. This discrepancy is likely a result of the increased precision of the measurements in this analysis. Whereas the DES analysis was focused on 2\% precision, our measurements had error bars on the order of 0.1\%. Nevertheless, \citet{2008.05991} was performed in configuration space so we present a more detailed comparison in \S\ref{sec:real_space_results}.

\begin{figure*}
\centering
\includegraphics[width=\linewidth]{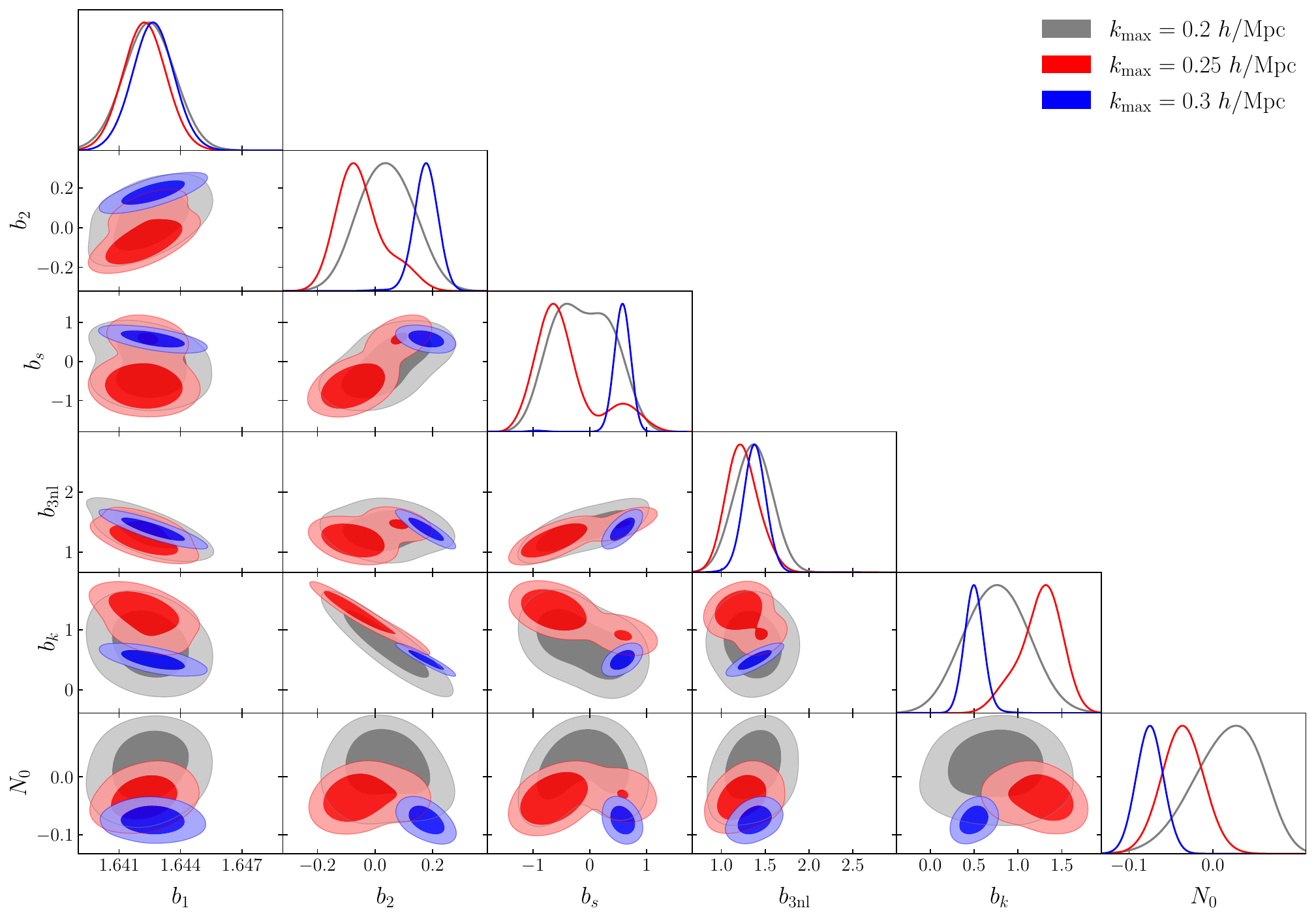}
\caption{Bias parameter posterior distribution when fitting the fiducial model to the ratios $P_{\rm gg}/P_{\rm mm}$ and $P_{\rm gm}/P_{\rm mm}$ for the $m_r<23$ galaxy sample at three scale cuts. }
\label{fig:Fourier_triangle}
\end{figure*}

\subsubsection{Comparison of r and i band selected samples}

In any realistic galaxy survey, the survey depth and hence the number density of detected objects will vary across the footprint. To a large extent this can be accounted for by using a magnitude limited sample, where the magnitude cut is chosen at sufficiently high signal-to-noise to ensure that the resulting catalog is complete across the footprint. Nevertheless, a magnitude cut in different bands will result in a slightly different sample selection. In this section we investigate the impact of this effect on our bias parameter estimates, as well as the dependence of our bias parameters on the scale cut.

Figure \ref{fig:bias_param_kmax} shows the bias parameter estimates as a function of ${k_{\rm max}}$ for our fiducial model fits for all six galaxy samples. The linear bias term, $b_1$, is stable for all galaxy samples at all scale cuts and is constrained with very high precision. The remaining bias parameters are constrained with significantly less precision and vary appreciably between scale cuts. Note that the dramatic shifts in several of the non-linear bias parameter estimates around $k_{\rm max}=0.25 \ h/{\rm Mpc}$ are a result of the multimodal likelihood surface discussed in \S\ref{sec:bimod}.

When comparing between the $r$ and $i-$band magnitude cuts, we find that the $b_1$ estimates between our matched samples differ by nearly $10 \sigma$ even though the approximate $b_1$ values are consistent to within 0.5\%. On the other hand, all higher-order bias parameters, with the exception of $b_2$ for the $m_r<22$ sample, agree to within $1 \sigma$ between corresponding magnitude limited galaxy samples suggesting these samples overlap considerably. While we have the sensitivity to detect these subtle differences in our analysis, they are unlikely to matter for any upcoming galaxy surveys.

Finally, Figure \ref{fig:bias_param_kmax} shows that our fiducial model estimates for $b_{\rm s}$ and $b_{\rm 3nl}$ deviate from their co-evolution values, particularly at large $k_{\rm max}$. These deviations are consistent with our model comparison findings in \S\ref{subsec:chi2_comparison} in which the fiducial model often provided significantly better fits to the data than the co-evolution models.

\begin{figure}[!t]
\includegraphics[width=\linewidth]{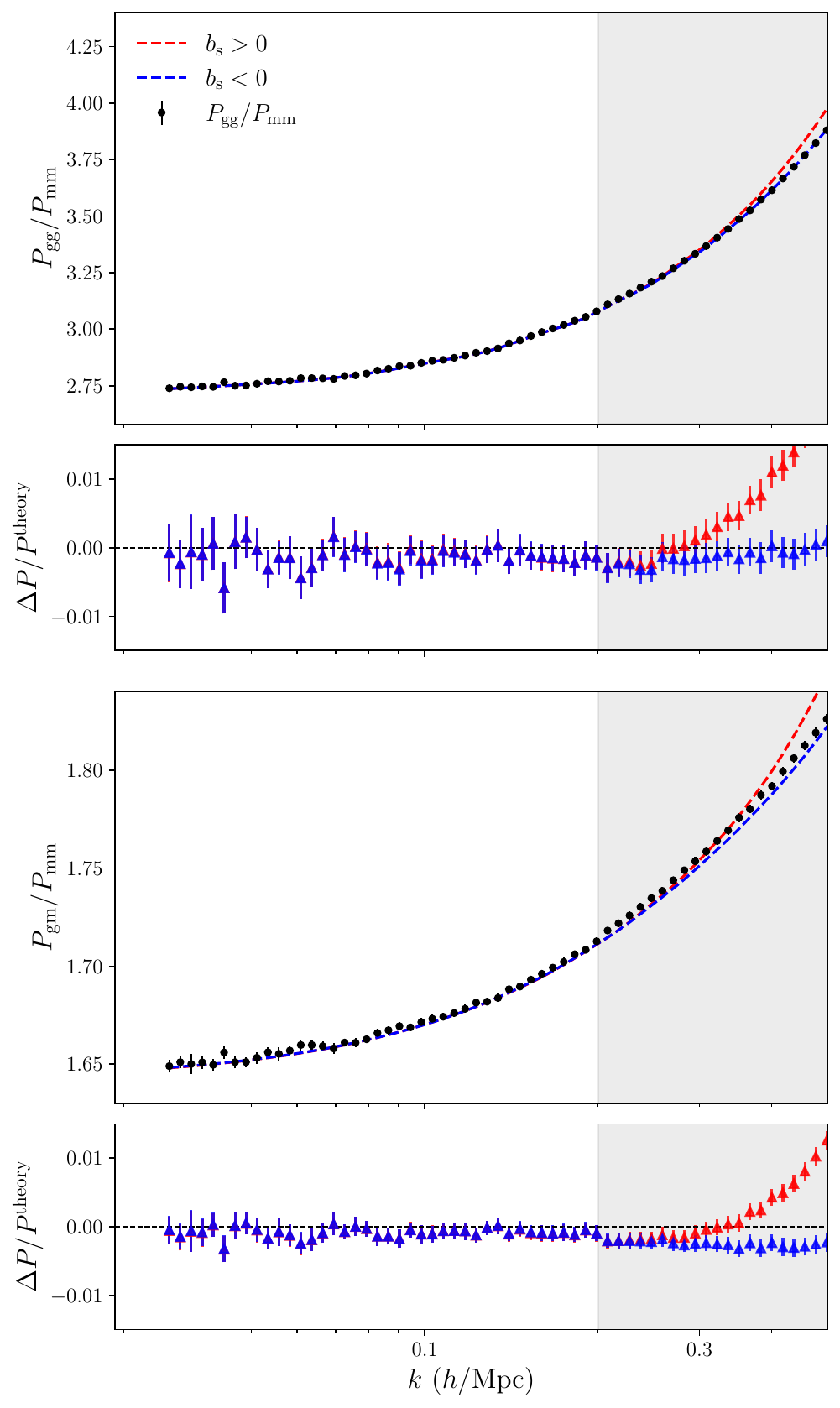}
\caption{$P_{\rm gg}/P_{\rm mm}$ and $P_{\rm gm}/P_{\rm mm}$ joint fit results for the $m_r<23$ galaxy sample at $k_{\rm max}=0.25 \ h/{\rm Mpc}$ using the best fit bias parameters derived from splitting the bimodal $b_s$ posterior.}
\label{fig:multimodal_fits}
\end{figure}

\subsubsection{Fit bimodality}
\label{sec:bimod}
Several of our Fourier space fiducial model bias parameter posteriors are bimodal for $k_{\rm max}\leq 0.25 \ h/$Mpc. Although a bimodal bias parameter distribution is not physically motivated, our constraints on higher-order bias parameters are $\mathcal{O}(1)$ and enter the model non-linearly, hence these parameters can generate a multimodal likelihood surface. In this section we investigate this multimodality.

Figure \ref{fig:Fourier_triangle} shows the bias parameter posterior distributions derived from the chains of the fiducial model fits for $m_r<23$ galaxies at three scale cuts using \texttt{GetDist} \cite{getdist}. The marginalized posteriors are reconstructed via kernel density smoothing of the chains which contain 9673, 10181, and 10651 samples for $k_{\rm max}=0.25, \ 0.30,$ and $0.35 \ h/{\rm Mpc}$, respectively, using the default settings described in \citet{getdist}. We check for convergence using an evidence tolerance of 0.1. At $k_{\rm max}=0.25 \ h/{\rm Mpc}$ the marginal posterior of $b_s$ is bimodal and the remaining bias parameters are relatively unconstrained compared to their estimates from fits assuming larger scale cuts. As we include more small scale information, the data favor $b_s$ values situated between the two modes with the bimodality disappearing for scale cuts beyond $k_{\rm max}=0.35 \ h/{\rm Mpc}.$ Note that the large error bars and occasional jumps in the maximum likelihood estimates of $b_s$ shown in Figure \ref{fig:bias_param_kmax} are a direct consequence of this bimodality.

\begin{figure}[!t]
\includegraphics[width=\linewidth]{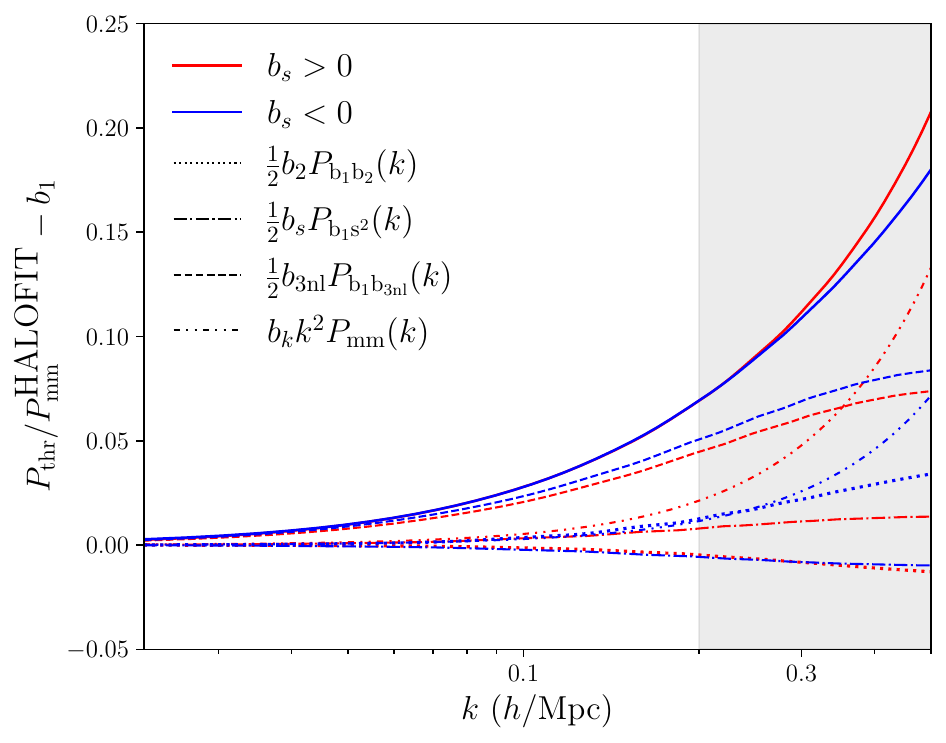}
\caption{Contributions of the PT terms in Eq. \ref{eq:P_tm} to $P_{\rm gm}/P_{\rm mm}$ using the positive and negative $b_s$ best fit bias parameters. The solid lines represent the sum of all non-linear contributions for the fiducial model.}
\label{fig:pt_kernel}
\end{figure}
\begingroup
\setlength{\tabcolsep}{5pt} 
\renewcommand{\arraystretch}{1.2}
\begin{table}[!htbp]
    \centering
     \begin{tabular}{| c | c|  c| c| c|  c|c|} 
     \hline
      $b_1$  & $b_2$ & $b_s$ & $b_{\rm 3nl}$ & $b_{\rm k}$ & $N_0$ &  $\chi^2/{\rm d.o.f}$ \\ [0.5ex] 
    \hline
    1.643 & 0.11 & 0.37 & 1.46 & 0.57 & 0.02& 0.478\\
     \hline
     1.642 & $-0.04$ & $-0.52$ & 1.28 & 1.05 & 0.02  & 0.475\\
     \hline
    \end{tabular}
    \caption{Maximum likelihood bias parameters for the fiducial model fits to the $m_r<23$ galaxy sample assuming
    $k_{\rm max} = 0.25 h/{\rm Mpc}$ for positive and negative values of $b_s$.}
    \label{tab:b_param_multimodal} 
\end{table}
\endgroup

\begin{figure*}
\centering
\includegraphics[width=\linewidth]{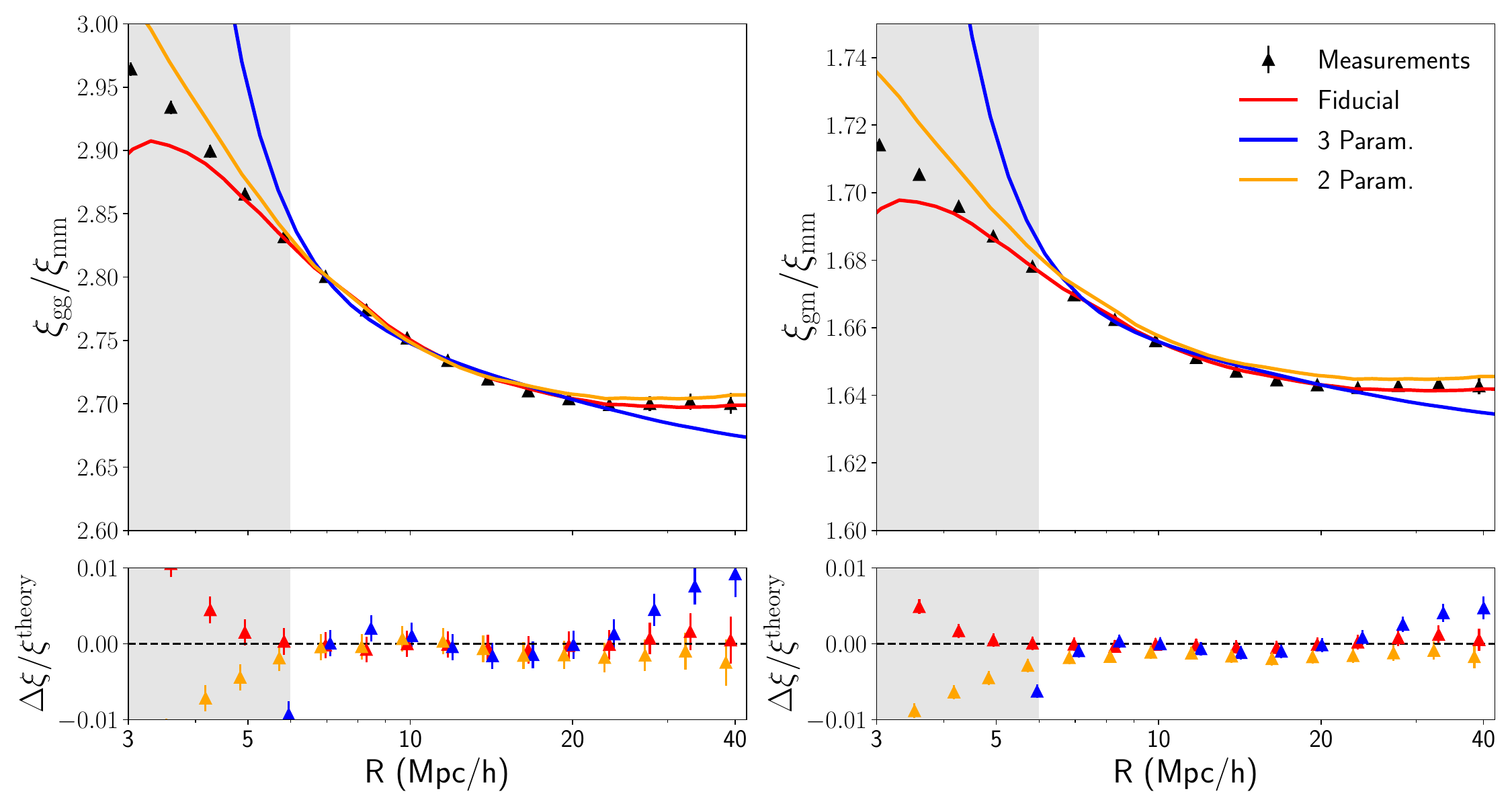}
\caption{Measured and bestfit configuration space correlations functions for the $m_r<23$ galaxy sample with $r_{\rm min}=6$ Mpc/$h$. The residuals ($\Delta \xi/\xi^{\rm theory}$) are shown in the bottom panels.
}
\label{fig:ratio_fits_real}
\end{figure*}
In order to compare the relative significance of the two modes at $k_{\rm max}=0.25 h/{\rm Mpc}$, we split the bias parameter sample into two chains based on the sign of $b_{s}$ and analyze the maximum likelihood parameter within each chain. In Table \ref{tab:b_param_multimodal} we list these parameters and their associated reduced $\chi^2$. Whereas the $b_1$ values are consistent between the two chains, the higher order bias parameter estimates disagree considerably. Nevertheless, the two sets of bias parameters yield comparable reduced $\chi^2$, hence both modes provide similar fits to the data given our likelihood, covariance, and scale cut. This is illustrated in Figure \ref{fig:multimodal_fits} which shows that the predictions of $P_{\rm gg}/P_{\rm mm}$ and $P_{\rm gm}/P_{\rm mm}$ for the two sets of bias parameters are essentially indistinguishable below the scale cut.

The observation that drastically different combinations of non-linear bias parameters can produce similar fiducial model predictions up to a fixed scale cut can be understood by analyzing the individual contributions of the terms in our fiducial model. In Figure \ref{fig:pt_kernel} we show the non-linear contributions to $P_{\rm gm}/P_{\rm mm}$ for the fiducial model using the bias parameters from Table \ref{tab:b_param_multimodal}. The solid lines represent the sum of all non-linear contributions to the galaxy-matter power spectrum in our fiducial model and are nearly identical for the two bias parameters within the scale cut. The dashed and dotted lines denote individual contributions to the $P_{\rm gm}/P_{\rm mm}$ model defined in Eq. \ref{eq:P_tm} from the various PT terms and can vary significantly between the modes. For example, the large scale contributions from the $b_2$ and $b_s$ terms in the fiducial model for the two modes effectively switch at low wavenumbers, yielding a comparable sum from different bias parameters. Similar deviations occur in the shapes of the remaining non-linear contributions with the $b_{\rm 3nl}$ ($b_k$) term providing a greater (lesser) contribution for the positive $b_{\rm s}$ mode (red) than the negative $b_{\rm s}$ mode (blue). These differences cancel at large scales when all PT terms are summed over.

\begin{figure}[!htbp]
\centering
\includegraphics[width=\linewidth]{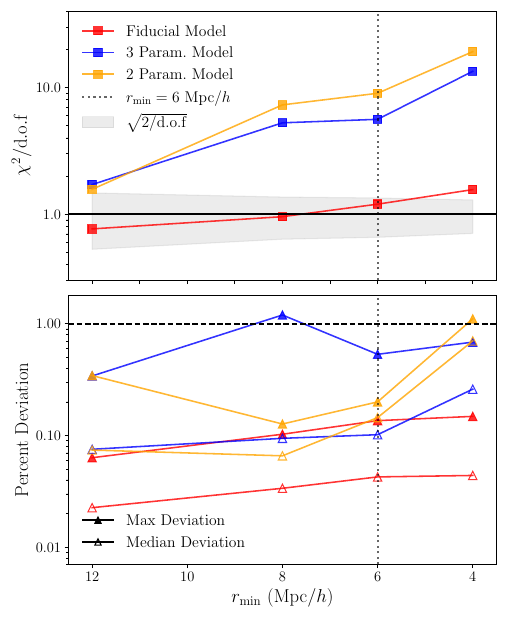}
\caption{\textit{Top:} Reduced $\chi^2$ as a function of $r_{\rm min}$ for the model fits in configuration space for $m_r<23$ galaxies. \textit{Bottom:} Maximum and median percent deviation for these model fits.}
\label{fig:chi2_real_space}
\end{figure}

\begin{figure*}
\centering
\includegraphics[width=\linewidth]{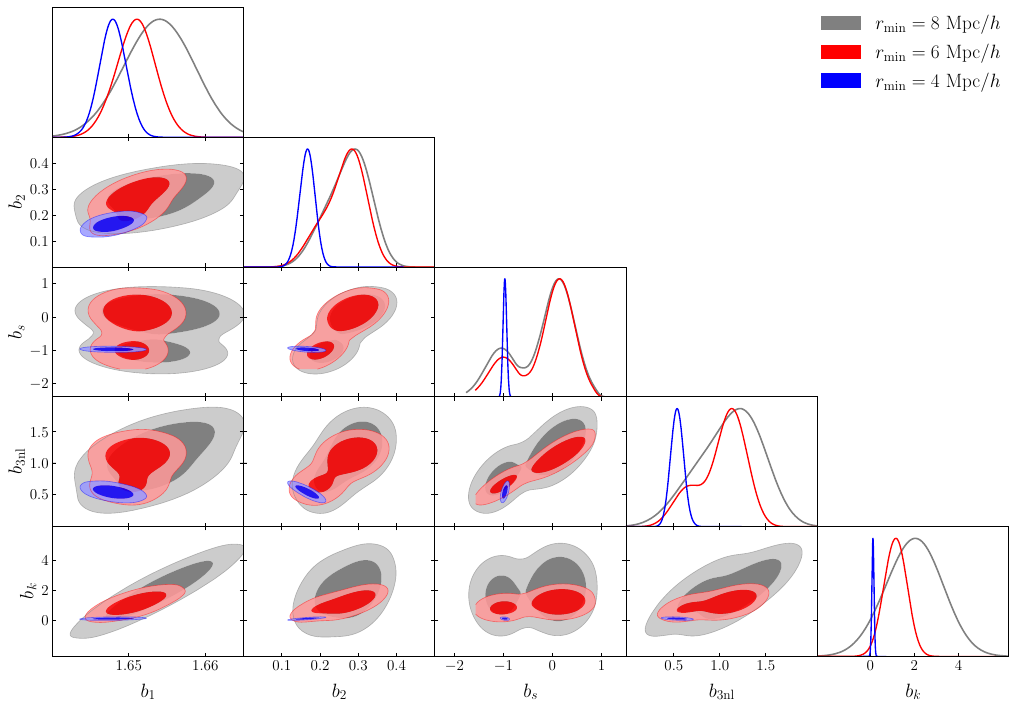}
\caption{Bias parameter posterior distribution when fitting the fiducial model to the ratios $\xi_{\rm gg}/\xi_{\rm mm}$ and $\xi_{\rm gm}/\xi_{\rm mm}$ for the $m_r<23$ galaxy sample at three different scale cuts.}
\label{fig:real_triangle}
\end{figure*}

In summary, cancellations between contributions of the non-linear PT terms in our fiducial model at sufficiently large scales can lead to theoretical predictions from disparate bias parameters that are indistinguishable within our error bars and hence yield bimodal bias parameter posteriors. A similar bimodality arises when fitting the configuration space data vectors $\xi_{\rm gg}/\xi_{\rm mm}$ and $\xi_{\rm gm}/\xi_{\rm mm}$ for the $m_r<23$ galaxy sample as shown in Figure \ref{fig:real_triangle}.

\subsection{Correlation function results}\label{sec:real_space_results}

In Figure \ref{fig:ratio_fits_real} we show the results from fitting the correlation function ratios $\xi_{\rm gg}/\xi_{\rm mm}$ and $\xi_{\rm gm}/\xi_{\rm mm}$ simultaneously with our three bias models. These model fits are for the $m_r<23$ galaxy sample and assume a scale cut of $r_{\rm min}=6 \  {\rm{Mpc}}/h$. We find an excellent fit with our fiducial model with five free parameters, which fits the correlation function measurements to within 0.2\% and has a reduced $\chi^2$ of 1.2. Similar to the Fourier space results for this galaxy sample, we find that the 2-parameter and 3-parameter models provide a reasonably good fit to the data vector, where the residuals are consistent with zero within a percent.

\begin{figure*}
\centering
\includegraphics[width=\linewidth]{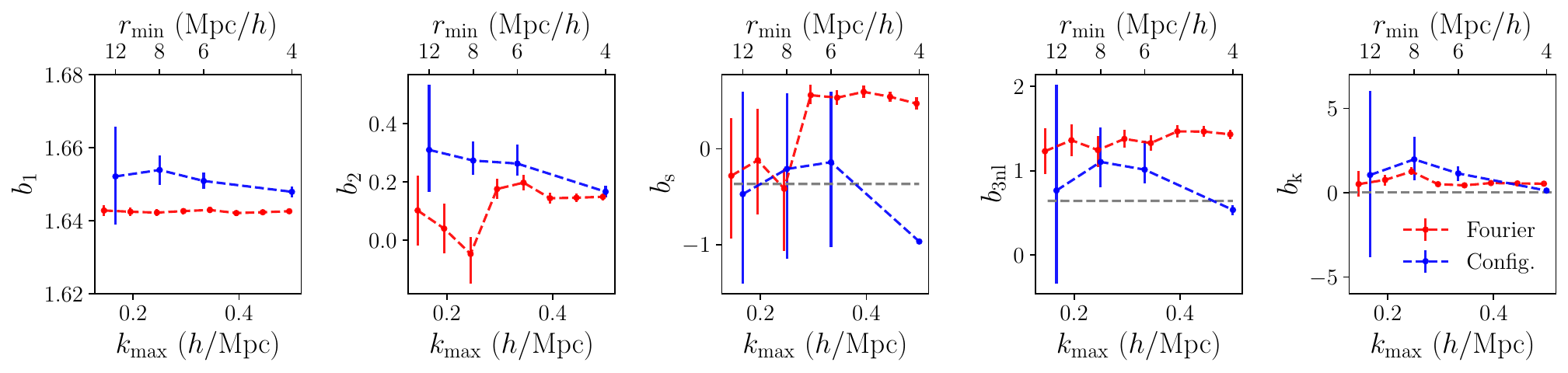}
\caption{Comparison of bias parameter estimates obtained from fitting in Fourier space and configuration space for the $m_r<23$ sample assuming scale cuts are related as $r_{\rm min}=2/k_{\rm max}$.}
\label{fig:Fourier_config_comp}
\end{figure*}

In Figure \ref{fig:chi2_real_space} we show the reduced $\chi^2$, maximum percent deviation, and median percent deviation for the configuration space model fits to the $m_r<23$ galaxy sample for four scale cuts ranging from $r_{\rm min}=4 \ {\rm Mpc}/h$ to $r_{\rm min}=12 \ {\rm Mpc}/h.$ Similar to the Fourier space results, the configuration space fiducial model fits have significantly lower reduced $\chi^2$ than those of the two and three parameter models across all scale cuts. However, unlike in Fourier space, the two and three parameter models yield comparable reduced $\chi^2$ statistics across all scale cuts when fitting this galaxy sample in configuration space. The configuration space fits also tend to have smaller reduced $\chi^2$ values than the Fourier space fits for the same galaxy sample. In particular, the reduced $\chi^2$ of the configuration space fiducial model fits is consistent with unity for all scale cuts analyzed. 

Similarly, the fiducial model demonstrates increased performance over the two and three parameter models when evaluated using the maximum and median percent deviation statistics. In particular, the configuration space fiducial model fits for the $m_r<23$ galaxy sample have maximum percent deviation below 0.15\% and median percent deviation below 0.05\% across all scale cuts. The maximum percent deviation of the two and three parameter models ranges from over 0.2\% to 1\%. We also find that the maximum percent deviation of the 2-parameter model fits is approximately less than or equal to that of the 3-parameter model fits at all except the most aggressive scale cut ($r_{\rm min}=4 \ {\rm Mpc}/h)$. This is consistent with the observation that the 3-parameter model provides a poorer fit to the large scale configuration space measurements in order to fit the data points between 10 and 20 ${\rm Mpc}/h$ which have tighter error bars as shown in the bottom panels of Figure \ref{fig:ratio_fits_real}. The 3-parameter model's inability to simultaneously describe both large and small scale information is a consequence of the fact that the non-linear PT terms can have an $\approx0.5\%$ contribution in the linear regime as discussed in Appendix \ref{Appendix:linear_bias}.

In Figure \ref{fig:real_triangle} we plot the bias parameter posteriors derived from the chains of the configuration space fiducial model fits for the $m_r<23$ galaxy sample at three scale cuts. As in Fourier space, we find that the marginalized posterior of $b_s$ is bimodal when assuming conservative scale cuts. This bimodality disappears with the addition of more small scale information; however, unlike the Fourier space results, the configuration space fiducial model fits favor the mode corresponding to negative values of $b_s$. As discussed in \S\ref{sec:bimod} this bimodality is a result of cancellations between the non-linear terms in our fiducial model for various combinations of bias parameters at large scales.

Our results differ from those of \citet{2008.05991} who found that the 2-parameter was sufficient to describe the 3D correlation function ratios $\xi_{\rm gg}/\xi_{\rm mm}$ and $\xi_{\rm gm}/\xi_{\rm mm}$ in DES-like mock galaxy catalogs from the \texttt{MICE} \cite{MICE:1, MICE:2, MICE:3} simulation. The insufficiency of the 2-parameter model in our analysis is likely a result of the increased statistical precision of our correlation function measurements. In particular,  \citet{2008.05991} found that the 2-parameter model fit the \texttt{MICE} correlation function measurement within 2\% up to 4 Mpc/$h$ which is similar to the percent level fits of the 2-parameter model achieved in this analysis.

\subsection{Comparison of Fourier and configuration space analyses}\label{sec:Fourier_config_comp}

In this section we compare bias parameter estimates between our Fourier space fits ($P_{\rm gg}/P_{\rm mm}$ and $P_{\rm gm}/P_{\rm mm}$) and configuration space fits ($\xi_{\rm gg}/\xi_{\rm mm}$ and $\xi_{\rm gm}/\xi_{\rm mm}$) for the $m_r<23$ galaxy sample. In Figure \ref{fig:Fourier_config_comp} we plot the bias parameters for our fiducial model as a function of $k_{\rm max}$ for Fourier space fits and $r_{\rm min}$ for configuration space fits assuming an conversion of the form $r_{\rm min}=2/k_{\rm max}$. Given that there is no exact correspondence between Fourier and configuration space scale cuts, we chose this approximate relation because it conveniently relates the range of scale cuts used in this analysis.

We find that the linear bias parameter, $b_1$, agrees to better than one percent between the Fourier space and configuration space fits across all scale cuts; however, the configuration space $b_1$ estimates are consistently higher than the estimates from the Fourier space fits. This tension in the linear bias parameter estimates is sourced by the behavior of the higher order PT kernels at the largest scales of our measurements, i.e. $0.035$ $h$/Mpc in Fourier space and 40 Mpc/$h$ in configuration space. At these scales, the non-linear terms in our fiducial model can still have a greater than percent level contribution to the power spectrum and correlation function theory predictions, and hence affect our constraints on $b_1$ which are derived from sub-percent measurements. We discuss this affect further in Appendix \ref{Appendix:linear_bias}.

The Fourier space fits also place tighter constraints on the bias parameters, particularly $b_1$. This is natural because $b_1$ is most strongly constrained on the largest scales, but our correlation function measurements only extend up to 40 Mpc/$h$. Furthermore, at large scales, the power spectra error bars are largely uncorrelated, unlike those of the correlation function. At smaller scales the signal-to-noise is greater, but the signal is also degenerate with that of the higher order bias parameters.  

\begin{figure}[!t]
\centering
\includegraphics[width=\linewidth]{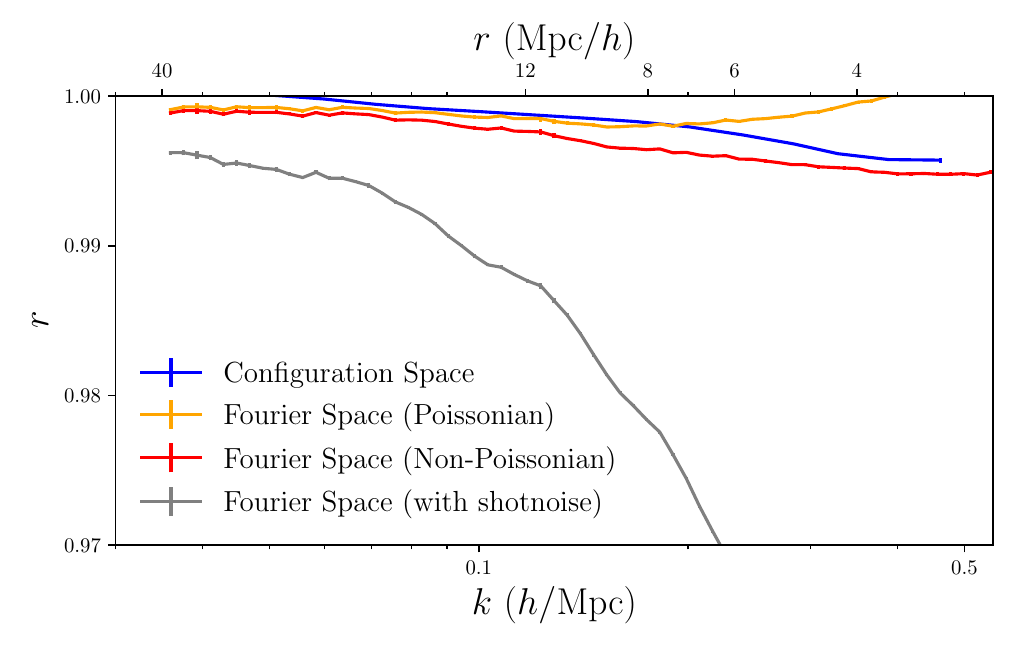}
\caption{Cross correlation coefficients in Fourier space and configuration space for the $m_r<23$ galaxy sample. The Fourier space cross correlation coefficient is shown after subtracting Poissonian shot-noise (orange), after subtracting the best fit non-Poissonian contribution for the fiducial model fit at $k_{\rm max}$=0.35 $h/{\rm Mpc}$ (red), and without subtracting any shot-noise (gray).}
\label{fig:cross_cor}
\end{figure}

The remaining bias parameters are determined with significantly less precision than $b_1$ in both Fourier and configuration space. We find significant tension in the $b_2$ estimates across all scale cuts; however, thae remaining bias parameters are relatively consistent between the two approaches below $k_{\rm max}=0.35 \ h/{\rm Mpc}$ and above $r_{\rm min}=6$ Mpc/$h.$ Finally, at $r_{\rm min}=4$ Mpc/$h$, the non-linear bias parameters disagree considerably between the two approaches. This is a result of the configuration space fits favoring the mode corresponding to negative values of $b_s$, as opposed to the positive mode selected in Fourier space as discussed in \S\ref{sec:bimod}. 

In Figure \ref{fig:cross_cor} we show the Fourier space and configuration space measured cross correlation coefficients $r(k)=P_{\rm gm}/\sqrt{P_{\rm mm}P_{\rm gg}}$ and $r(r)=\xi_{\rm gm}/\sqrt{\xi_{\rm mm}\xi_{\rm gg}}$. The Fourier space measurement are shown after subtracting Poissonian shot-noise, as well as after subtracting the non-Poissonian shot-noise inferred from the fiducial model fit at $k_{\rm max}=0.35 \ h/{\rm Mpc}.$ The Fourier and configuration space axes are arbitrarily related by $r=1.4/k$ because it provides a convenient comparison of scales of our measurements. It is difficult to directly compare $r$ between the configuration and Fourier space because of the non-trivial relationship between $r$ and $k$, as well as the dependence of the Fourier space measurement on the shot-noise estimate. Nevertheless, we find that all cross-correlation coefficients are remarkably consistent with unity well beyond the linear regime.

Differences in bias parameter estimates obtained from the two approaches can be expected given that configuration space and Fourier space fundamentally encode different information under fixed scale cuts. Moreover, higher order contributions to the power spectrum and correlation function data vectors that are not accounted for in the bias model can be fit differently between the two spaces, hence leading to shifts in bias parameters. Nevertheless, several of our derived bias values are inconsistent at high statistical significance. Since the aforementioned processes should have only secondary effects on the data vectors and theory modeling, our observed differences between the two approaches could be a result of systematic numerical effects (in either theory predictions or measurements) or fitting the noise features in the simulation. The tension between our two approaches suggests that it is difficult to draw conclusions from the bias parameter values, particularly the non-linear parameters. These parameters should likely be treated as nuisance parameters and marginalized over. 

Finally, we find that the configuration (Fourier) space results are unable to describe the Fourier (configuration) space measurements. In particular, using the best fit bias parameters from the configuration space fiducial model fits with the non-Poissonian shot-noise from the Fourier space fits to describe the Fourier space data vectors to $k_{\rm max}=0.35 \ h/{\rm Mpc}$ yields reduced $\chi^2$ values of 32.8, 168, 652, and 74.5 at $r_{\rm min}=4, 6, 8,$ and 12 ${\rm Mpc}/h$. This incompatibility is driven by the large $b_{\rm k}$ values of the configuration space fits which lead to overestimates of $P_{\rm gg}/P_{\rm mm}$ and $P_{\rm gm}/P_{\rm mm}.$

\begin{figure}[!tbp]
\centering
\includegraphics[width=\linewidth]{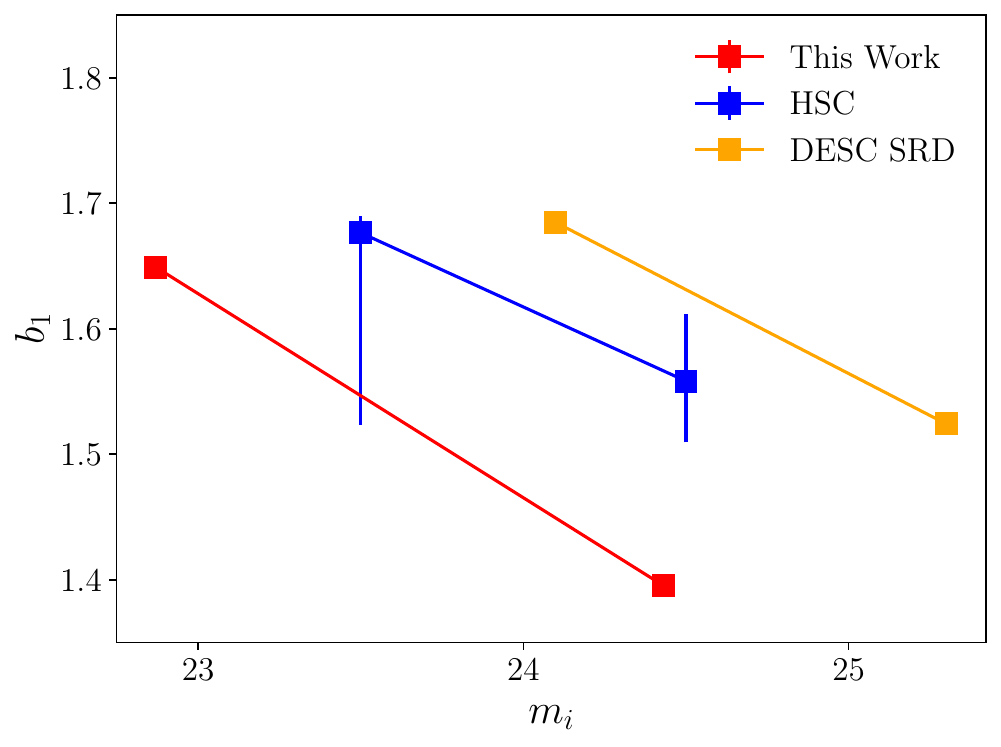}
\caption{Comparison of our linear bias parameter estimates with observations of HSC galaxy clustering \cite{1912.08209}, as well as the LSST DESC SRD \cite{1809.01669}.}
\label{fig:observational_comparison}
\end{figure}

\subsection{Comparison with measurements}

The linear bias parameter estimates from the fiducial model fits in our analysis are 1.394$\pm 0.001$, 1.649$\pm 0.001$ and 1.880$\pm 0.002$ for $m_i<24.43$, $m_i<22.87$ and $m_i<21.80$ samples, respectively. The LSST DESC Science Requirements Document (SRD) \cite{1809.01669} assumed the form
\begin{equation}
b_1(z) = \frac{b_1(z=0)}{g(z)},
\end{equation}
with $b_1(z=0)=0.95$ and 1.05 for $m_i<25.3$ and $m_i<24.1$, implying $b_1(z=1.0)=1.51$ and $1.68$ for those two galaxy samples. Analysis of HSC galaxy clustering in \citet{1912.08209} found bias values lower than those provided by the SRD, with $b_1(z=1.0)\sim1.60$ and $\sim1.55$ for $m_i<23.5$ and $m_i<24.5$. Our analysis of \CosmoDC galaxies gives $b_1$ values about 10\% lower than the HSC measurements. We present a comparison of these estimates in Figure \ref{fig:observational_comparison}.

\section{Conclusions}\label{sec:conclusions}

Effectively modeling non-linear galaxy bias is necessary to extract cosmological information from upcoming galaxy surveys. Large N-body simulations provide an invaluable avenue to rigorously test non-linear galaxy bias models. In this paper we validated a hybrid-PT bias model against galaxy samples from the \CosmoDC simulations. We measured and analyzed the two-point correlations between the galaxy and matter catalogs in both Fourier and configuration spaces. These measurements are obtained from a single 27 $({\rm Gpc}/h)^3$ box at redshift $z\sim 1$ and are some of the highest precision measurements to date. We list our main findings below:
  \begin{itemize}
      \item A five-parameter EFT-inspired hybrid perturbation theory model can effectively describe  non-linear galaxy bias in the 3D power spectrum measurements from \CosmoDC at the sub-percent level up to ${\rm k_{\rm max}}=0.4  \ h/{\rm Mpc}$ and with 2\% precision up to  ${\rm k_{\rm max}} = 0.9 \ h/{\rm Mpc}$ for all magnitude limited samples considered in this paper at $z \sim 1$. Furthermore, after fixing three of the higher order bias parameters to values obtained from theoretical considerations and letting only $b_1$ and $b_2$ vary, we are still able to fit the measurements with 2\% precision up to  ${\rm k_{\rm max}} = 0.5 \ h/{\rm Mpc}$.
      
      \item Bias parameter estimates obtained from distinct galaxy samples defined via $r$-band and $i$-band magnitude that are matched to give the same number of galaxies are essentially the same between the two samples. In particular, abundance matching \cite{1011.4964}, which implies that the ensemble properties of galaxies are ordered by some measure of their ``bigness'' (host halo mass, stellar mass, or flux in some band in our case) works to better than percent level in this analysis.
      
      \item The bias parameter likelihood surface for our fiducial model fits in Fourier space and configuration space is often multimodal when assuming relatively conservative scale cuts. This multimodality arises from degeneracies between the large scale contributions of the various PT kernels.
      
      \item Bias parameters estimated from fitting the power spectra in Fourier space exhibit moderate to significant tension with those obtained from fitting the configuration space correlation functions. In particular, although estimates of $b_1$ agree to within 1\% for the two methods, there is still considerable tension between the configuration space and Fourier space linear bias estimates given the precision of our measurements. The remaining bias parameters, with the exception of $b_2$, are consistent within $1\sigma$ between the two approaches until scale cuts of $k_{\rm max}=0.35 \ h/{\rm Mpc}$ and $r_{\rm min}=6 \ {\rm Mpc}/h$. These discrepancies likely arise from a combination of numerical effects in the measurement procedure, incompleteness of our bias model, and intrinsic differences between information contained in the power spectra and correlation functions when assuming fixed scale cuts. Since cosmological constraints come predominately from correctly determining the value of $b_1$, this should not be a major obstacle for most cosmological analysis. We leave pursuing a detailed understanding of  this discrepancy for future work.

      \item At the sub-percent precision of our measurements, we find significant departures of the non-linear parameters $b_{\rm s}$ and $b_{\rm 3nl}$ from their expected co-evolution values (when assuming a linear galaxy biasing in Lagrangian space). While we do expect departures from  co-evolution at some level, given the differences between Fourier and configuration space fits it is not clear whether these are real or systematic effects.
  \end{itemize}
  
  This study complements various recent studies aiming to study  galaxy bias models with high resolution simulations \cite{2008.05991, sugiyama_2020, Eggemeier_2020, Eggemeier_2021, Saito2014a, Angulo_2015, bella2018impact, Werner_2019, alex2020testing}. Unlike previous studies which focus either on Fourier or configuration space, in this study we perform the analysis using the two point functions in both spaces. As these two estimators are sensitive to different scales and hence differ in their sensitivity to various non-linearities, they provide a valuable check of the consistency of the cosmological and astrophysical conclusions from the data. Therefore a study like the one presented here would be useful in making model and analysis choices for future and current galaxy surveys. A majority of the previous studies used halos as the tracers of dark matter, whereas this analysis focused specifically on galaxy bias. We analyze both bright and faint galaxy samples at high redshift, where we expect future galaxy surveys to have peak sensitivity. Our findings suggest that these surveys should use the full five parameter model to describe sub-percent level measurements; however, the two and three parameter models are likely to provide sufficient descriptions at the 1-2\% level as shown in Figures \ref{fig:Fourier_space_reduced_chi2} and \ref{fig:chi2_real_space}.

  In this study, we have not included the effects of redshift space distortions or lensing magnifications on our measurements and model. These physical processes can be easily accounted for at linear level \cite{1911.11947}, but we leave a detailed study of the impact of their higher order contributions to a future study. The EFT-inspired model used in this work is ``complete'' in that it can describe any halo-distribution on sufficiently large scales. While not perfect, we find that for a realistic distribution of galaxies, it allows sub-percent descriptions to surprisingly large values of $k_{\rm max}$. Including baryonic and other non-gravitational processes that can affect galaxy formation will modify this picture. The next step is to confront this model with other models in the literature, such as Lagrangian perturbation theory \cite{Bouchet95, Matsubara08, Carlson13, Vlah15}, hybrid effective field theory \cite{Modi_2017,HEFT_DESY1}, and effective field theory of large scale structure \cite{Baumann_12,Senatore_15,Angulo_2015,DAmico_eft:2019}. It is hopeful that more than one theoretical framework will produce unbiased results with different and informative sensitivity to systematic effects.

\section*{Acknowledgements}

This paper has undergone internal review in the LSST Dark Energy Science Collaboration. The authors acknowledge feedback from the internal reviewers: Phil Bull, Joe DeRose, and Antonia Sierra Villarreal. 

DESC acknowledges ongoing support from the IN2P3 (France), the STFC 
(United Kingdom), and the DOE, NSF, and LSST Corporation (United States).  
DESC uses resources of the IN2P3 Computing Center 
(CC-IN2P3--Lyon/Villeurbanne - France) funded by the Centre National de la
Recherche Scientifique; the National Energy Research Scientific Computing
Center, a DOE Office of Science User Facility supported under Contract 
No.\ DE-AC02-05CH11231; STFC DiRAC HPC Facilities, funded by UK BIS National 
E-infrastructure capital grants; and the UK particle physics grid, supported
by the GridPP Collaboration.  This work was performed in part under DOE 
Contract DE-AC02-76SF00515.

BJ and SP are supported in part by the US Department of Energy Grant No. DE-SC0007901.

This work was supported in part by the U.S. Department of Energy, Office of Science, Office of Workforce Development for Teachers and Scientists (WDTS) under the Science Undergraduate Laboratory Internships Program (SULI). 

\appendix

\begin{figure}[!t]
\centering
\includegraphics[width=\linewidth]{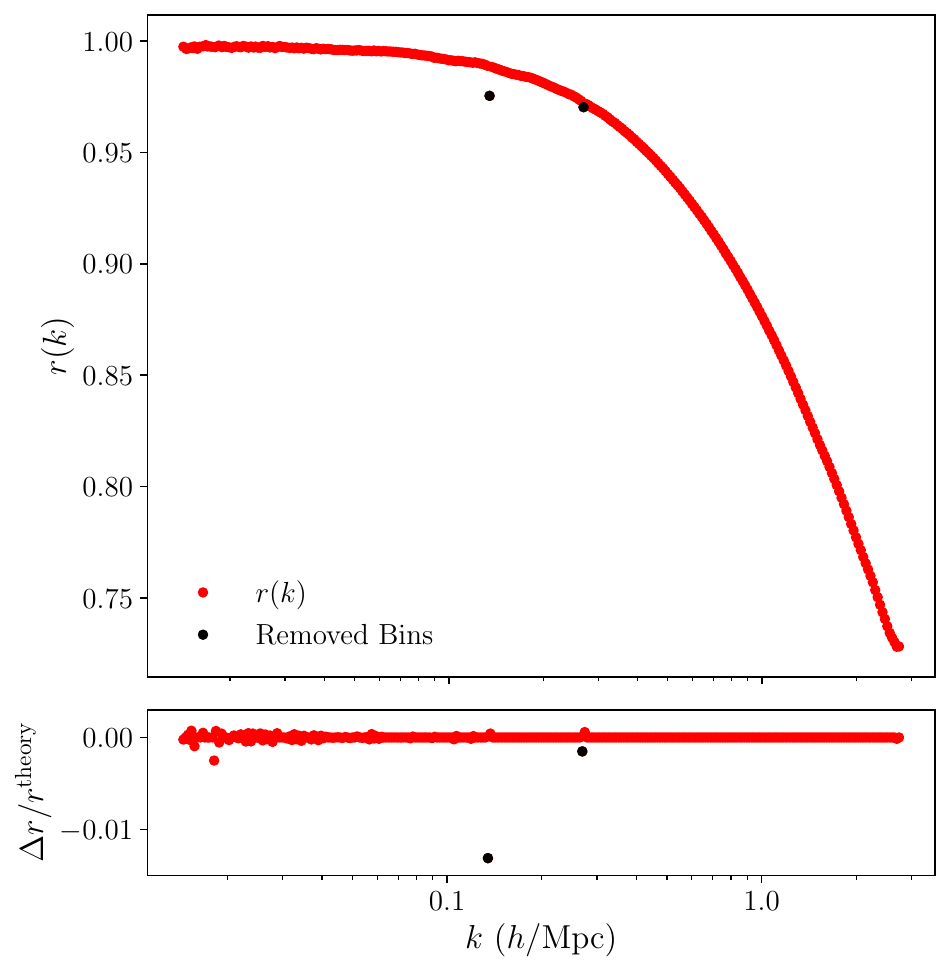}
\caption{\textit{Top:} Galaxy-matter cross correlation coefficient for the $m_r<24.5$ galaxy sample. The two black modes with outlying values of $r(k)$ resulting from excess power in $\rm{P}_{mm}$ are removed before rebinning the measured power spectra. \textit{Bottom:} Galaxy-matter cross correlation coefficient residuals calculated using a 3 point moving median about each bin center for the theory model, $r^{\rm theory}.$ }
\label{fig:excess_power_r_k}
\end{figure}

\begin{figure}
\centering
\includegraphics[width=\linewidth]{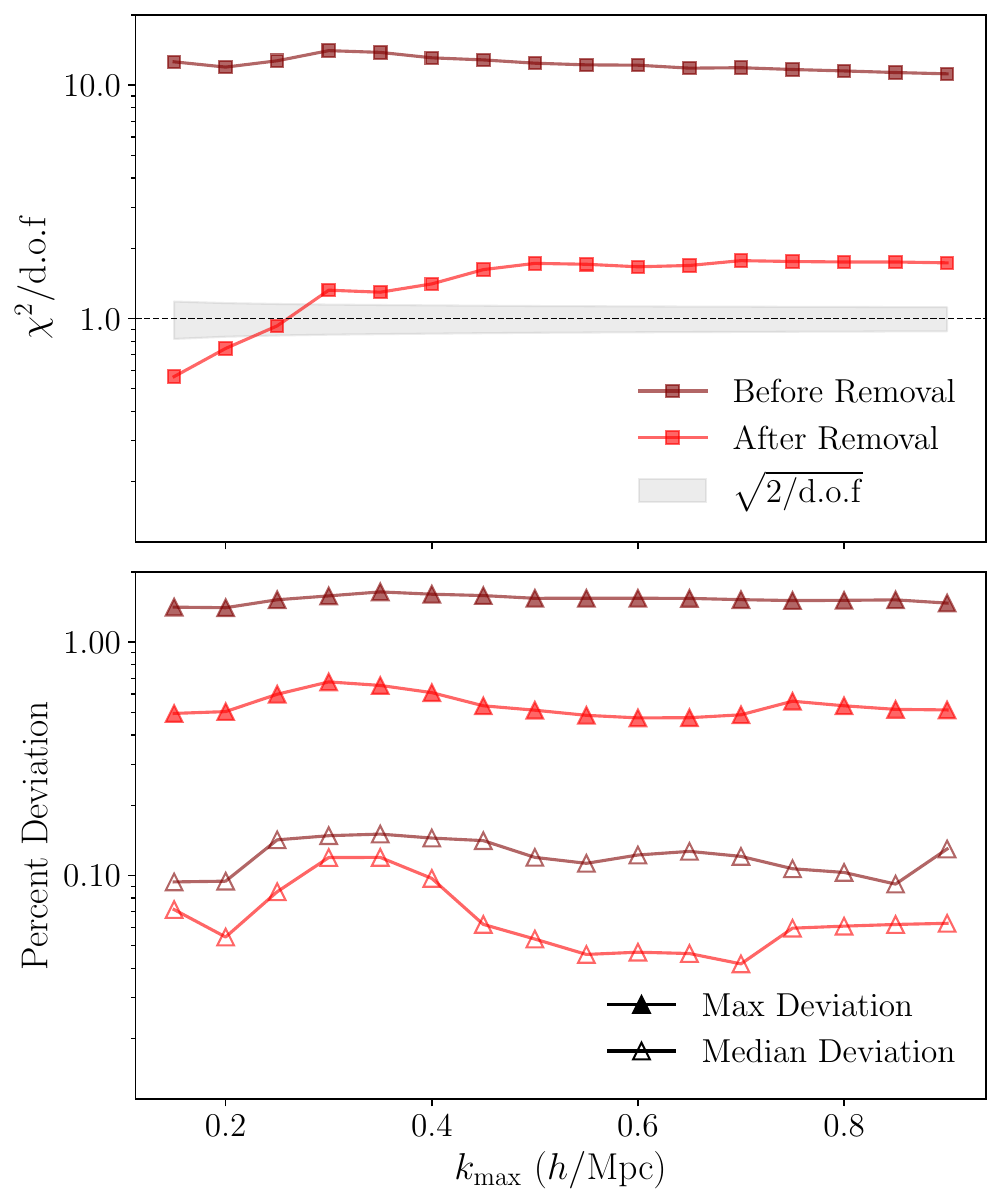}
\caption{\textit{Top:} Reduced $\chi^2$ as a function of $k_{\rm max}$ for the fiducial model fits in Fourier space using the $m_r<24.5$ galaxy sample with and without the outlying bins. \textit{Bottom: } Maximum and median percentage deviation for the fiducial model fits to the $m_r<24.5$ galaxy sample with and without the outlying bins.}
\label{fig:chi2_bin_remov}
\end{figure}

 \begin{figure*}
\centering
\includegraphics[width=\linewidth]{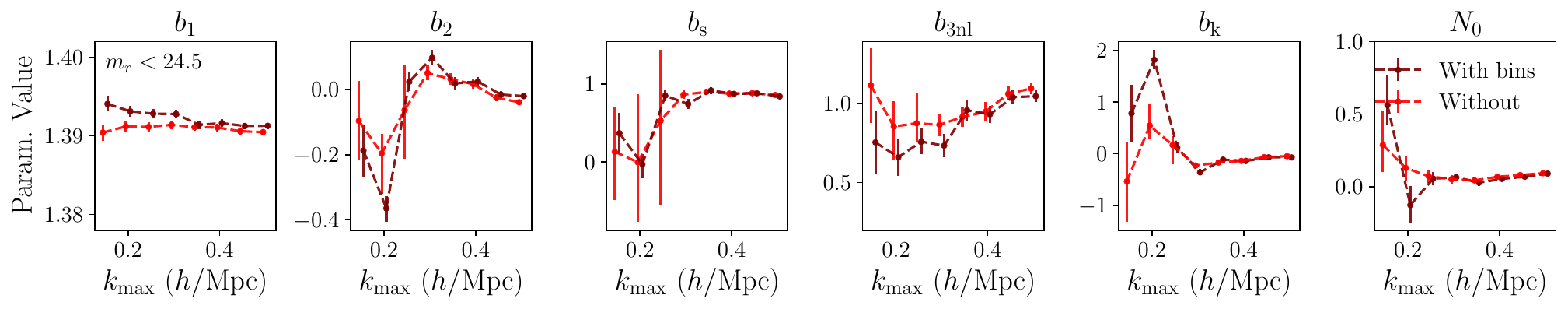}
\caption{Bias parameters as a function of $k_{\rm max}$ for the fiducial model fits in Fourier space using the $m_r<24.5$ galaxy sample with and without the outlying bins shown in Figure \ref{fig:excess_power_r_k}.}
\label{fig:param_kmax_bin_removal}
\end{figure*}

\section{Removal of excess power} \label{Appendix:Excess_Pow}
When analyzing the galaxy-matter cross correlation coefficient across all galaxy samples using 300 logarithmically spaced bins between $0.01$ and $2.8 \ h/{\rm Mpc}$ we found several modes containing significant outliers. In Figure \ref{fig:excess_power_r_k} we plot the cross correlation coefficient for the $m_r<24.5$ galaxy sample in 300 bins between 0.01 and 2.8 $h/\rm{Mpc}$. The two modes labeled in black represent significant outliers as can be seen in the bottom panel in which we plot the cross correlation coefficient residuals using a 3 point moving median. These outliers result from excesses in the matter power spectrum corresponding to wavenumbers where $k_x$ is equal to an integer multiple of $64$ regardless of our choice of binning. We treated the finding as a simulation artifact and removed these two bins from all of our full box power spectra measurements. Note that we do not manually remove the bin associated with the outlier at $k= \ 0.018  h/{\rm Mpc}$ as it is automatically excluded when we rebin using a minimum wavenumber of 0.035 $h/{\rm Mpc}.$

To understand the impact of removing these bins we also ran the fiducial model fitting procedure with these bins included. Leaving these bins shifts the corresponding measurement bins of $P_{\rm gg}/P_{\rm mm}$ and $P_{\rm gm}/P_{\rm mm}$ by approximately 0.5\% which is greater than the precision of our measurements. In effect, including these modes in our final data vector impacts our model fits.

Figure \ref{fig:param_kmax_bin_removal} shows the fiducial model bias parameter constraints before and after removing the outlying bins. The inferred bias parameters differ between the two samples, particularly at low $k_{\rm max}$ where the model fits are heavily constrained by the outlying bin located $k= 0.14 \ h/\rm{Mpc}$. Similarly, including these two bins in our analysis significantly degrades our goodness of fit statistics. In Figure \ref{fig:chi2_bin_remov} we plot the reduced $\chi^2$, maximum percentage deviation, and median percentage deviation as a function of $k_{\rm max}$ for the fiducial model fits of the $m_r<24.5$ galaxy sample with and without including the two bins with excess matter power. Removing these bins yields a significant drop in the reduced $\chi^2$ statistic, indicating that these bins impact the model fits appreciably. Including these modes leads to a considerable increase in the maximum percentage deviation as the model is unable to fit these outliers; nevertheless, the median percent deviation is relatively unaffected. These findings motivate our decision to treat the outliers as a systematic error and remove them from our analysis.

\section{Covariance matrix estimation} \label{Appendix:Covariance_Adjustment}
The Fourier space covariance matrices estimated using the subsample covariance of the ratios $P_{\rm gg}/P_{\rm mm}$ and $P_{\rm gm}/P_{\rm mm}$ calculated in 64 subboxes are essentially singular, hence leading to unstable model fits and bias parameter estimates. To address this, we artificially inflated the diagonal of all Fourier space covariance matrices by 5\%. In this section we describe the impact of this adjustment on our bias parameter estimates and model fits.

\begin{figure}[!b]
\centering
\includegraphics[width=\linewidth]{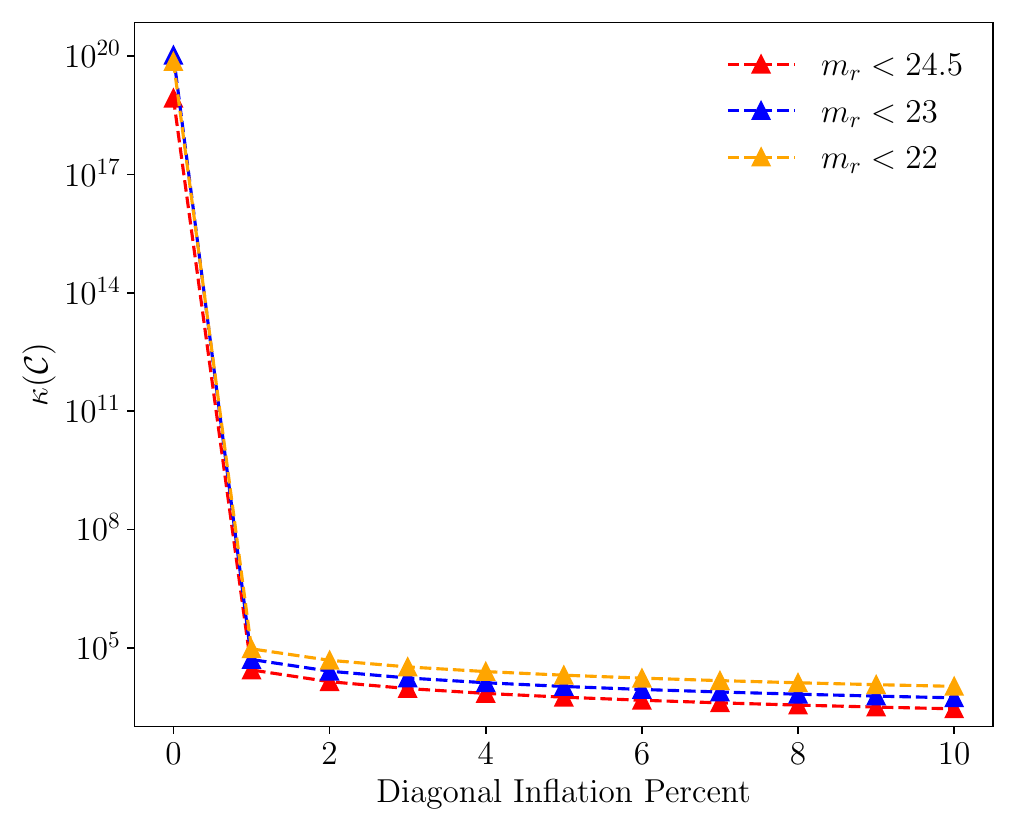}
\caption{Dependence of condition number of Fourier space covariance matrices for $r$-band galaxy samples on the diagonal inflation percentage of the covariance matrix assuming $k_{\rm max}= 0.5 \ h/{\rm Mpc}.$ }
\label{fig:condition_number}
\end{figure}

\begin{figure*}
\centering
\includegraphics[width=\linewidth]{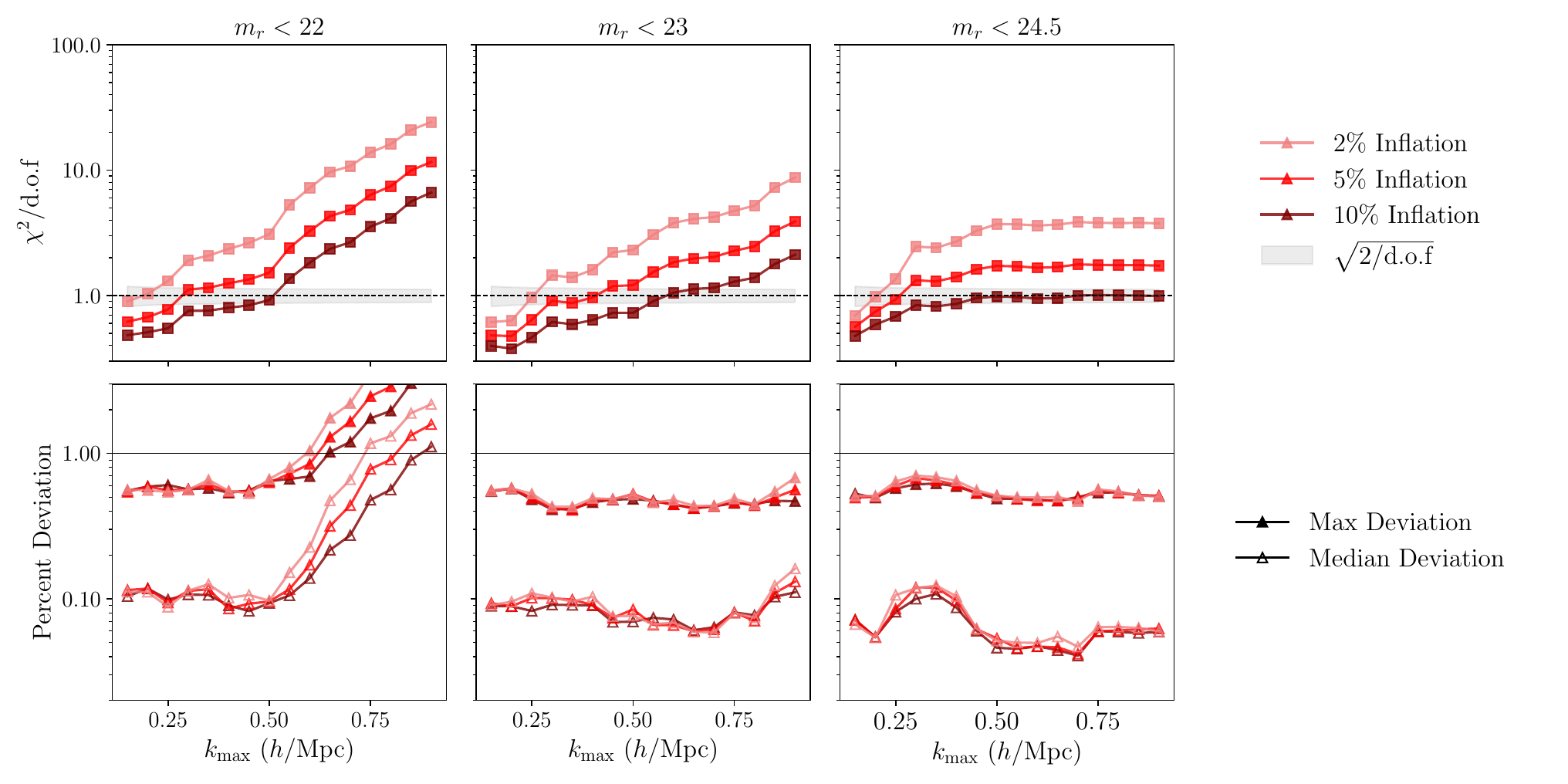}
\caption{\textit{Top: } Reduced $\chi^2$  as a function of $k_{\rm max}$ for the fiducial model fits to the three $r$-band galaxy samples after inflating the diagonal of the covariance matrix by 2\%, 5\%, and 10\%. \textit{Bottom: } Maximum (filled) and median (unfilled) percent deviation for these model fits.}
\label{fig:chi2_diag_inflate}
\end{figure*}

 \begin{figure*}
\centering
\includegraphics[width=\linewidth]{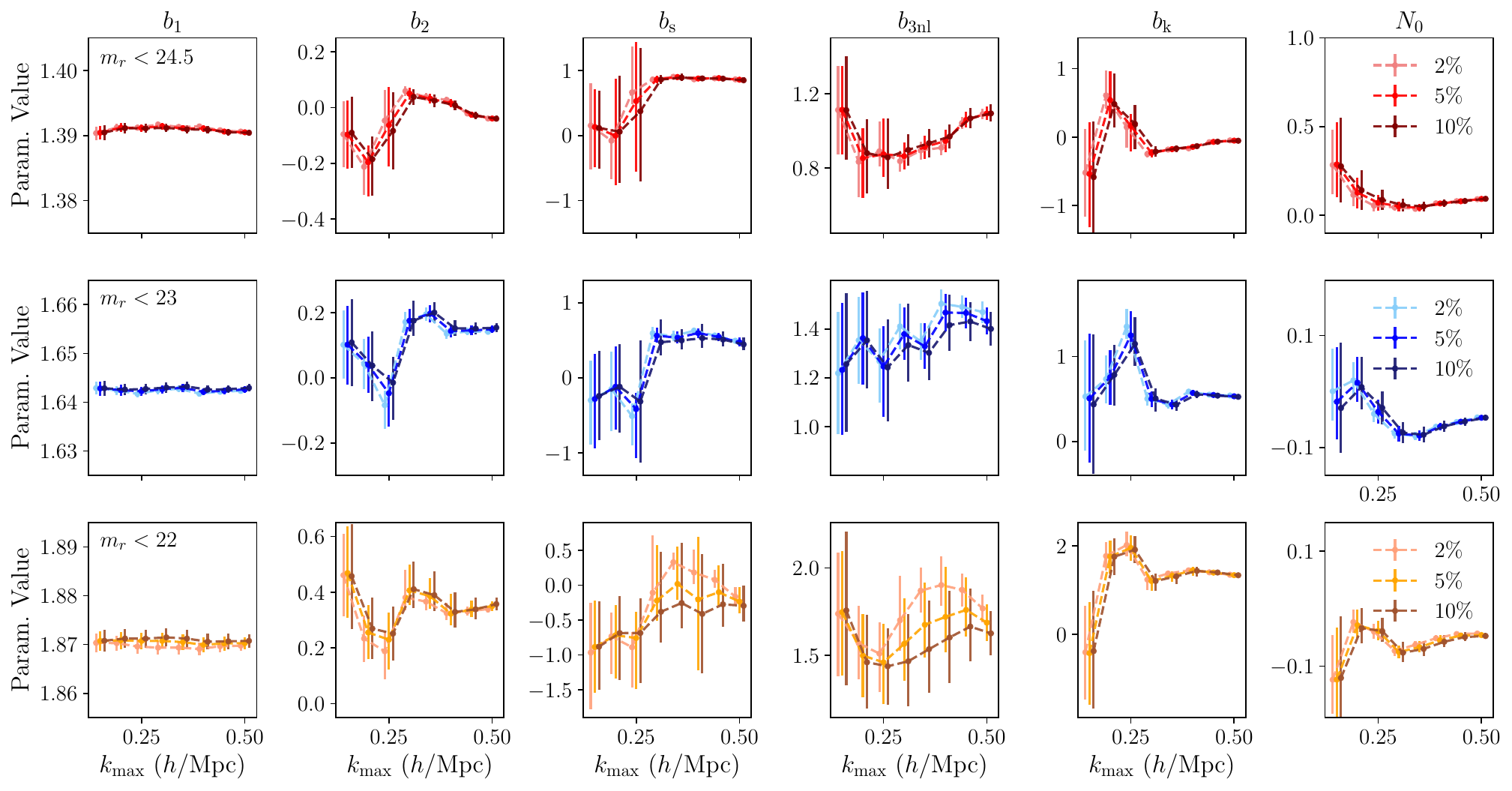}
\caption{Bias parameters as a function of $k_{\rm max}$ for the fiducial model fits in Fourier space across all $r$-band magnitude limited galaxy samples after inflating the diagonal of the covariance matrix by 2\%, 5\%, and 10\%.}
\label{fig:param_kmax_cov_inflate}
\end{figure*}

In Figure \ref{fig:condition_number} we plot the condition number, $\kappa(\mathcal{C})$, of the Fourier space covariance matrices for all $r$-band limited galaxy samples as a function of the percent increase in the covariance matrix diagonal assuming $k_{\rm max} = \ 0.5 h/{\rm Mpc}.$ We find that the condition number drops by over two orders of magnitude after increasing the diagonal of the covariance matrix by 1\%. With larger adjustments, the condition number continues to steadily fall. At 5\%, our selected inflation percentage, the condition numbers for all analyzed galaxy samples are of order $10^4.$

In Figure \ref{fig:chi2_diag_inflate} we plot the reduced $\chi^2$, maximum percentage deviation, and median percentage deviation as a function of $k_{\rm max}$ for all three $r$-band galaxy samples and all three perturbative bias models with the covariance matrices inflated by 2\%, 5\%, and 10\%. The $\chi^2/\rm{d.o.f}$ decreases appreciably as the diagonal is inflated; therefore, we need to introduce alternative metrics in order to reliably evaluate the goodness of fit of our models while maintaining stable fits. Unlike the reduced $\chi^2,$ the maximum and median percentage deviation statistics are relatively stable with respect to inflating the diagonal of the covariance matrix and the largest deviations are $\mathcal{O}(0.1)$. Therefore, our percent level claims remain unaltered by our covariance matrix adjustment.

In Figure \ref{fig:param_kmax_cov_inflate} we show the bias parameter estimates obtained from fitting the $r$-band limited galaxy samples to the fiducial model with a 2\%, 5\%, and 10\% inflation of the covariance matrix diagonal. We find extremely consistent agreement between essentially all bias parameters across all scale cuts regardless of the choice of diagonal inflation. There are slight deviations in the parameter estimates of $b_{\rm{s}}$ for the $m_r<22$ galaxy sample and $b_{\rm{3nl}}$ for the $m_r<24.5$ at large $k_{\rm max}$; however, these parameters are still consistent to within $1 \sigma.$

\begin{figure}[!tbp]
\centering
\includegraphics[width=\linewidth]{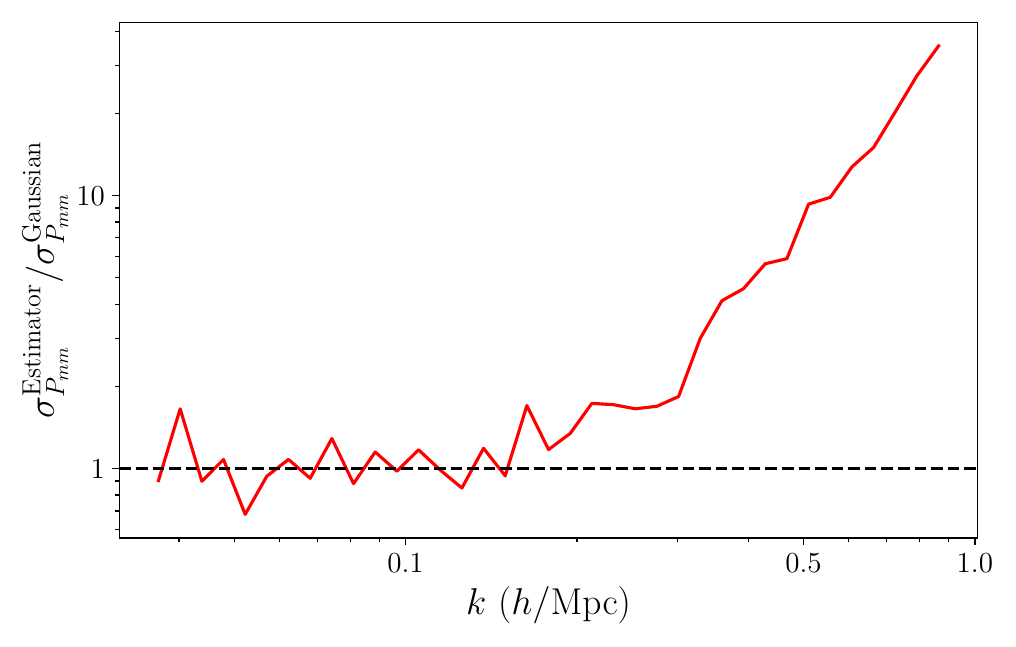}
\caption{Ratio of the standard deviation of $P_{\rm mm}$ computed using the Fourier space estimator to its Gaussian approximation.}
\label{fig:gauss_cov}
\end{figure}

Finally, in Figure \ref{fig:gauss_cov} we compare our estimated error on $P_{\rm mm}$ using the subsample covariance with a 5\% inflation of the main diagonal with the Gaussian error approximation given by $\sigma^{\rm Gaussian}_{P_{\rm mm}}=\sqrt{\frac{2P_{\rm mm}^2}{N_{\rm modes}}}$. We find agreement between the two methods up to $k \approx 0.15$ $h$/Mpc after which our errors are notably larger than the Gaussian estimate. This is consistent with the scales at which non-Gaussian errors are expected to become important at $z=1.01$ \cite{Blot_2015}.

 \begin{figure}[!t]
\centering
\includegraphics[width=\linewidth]{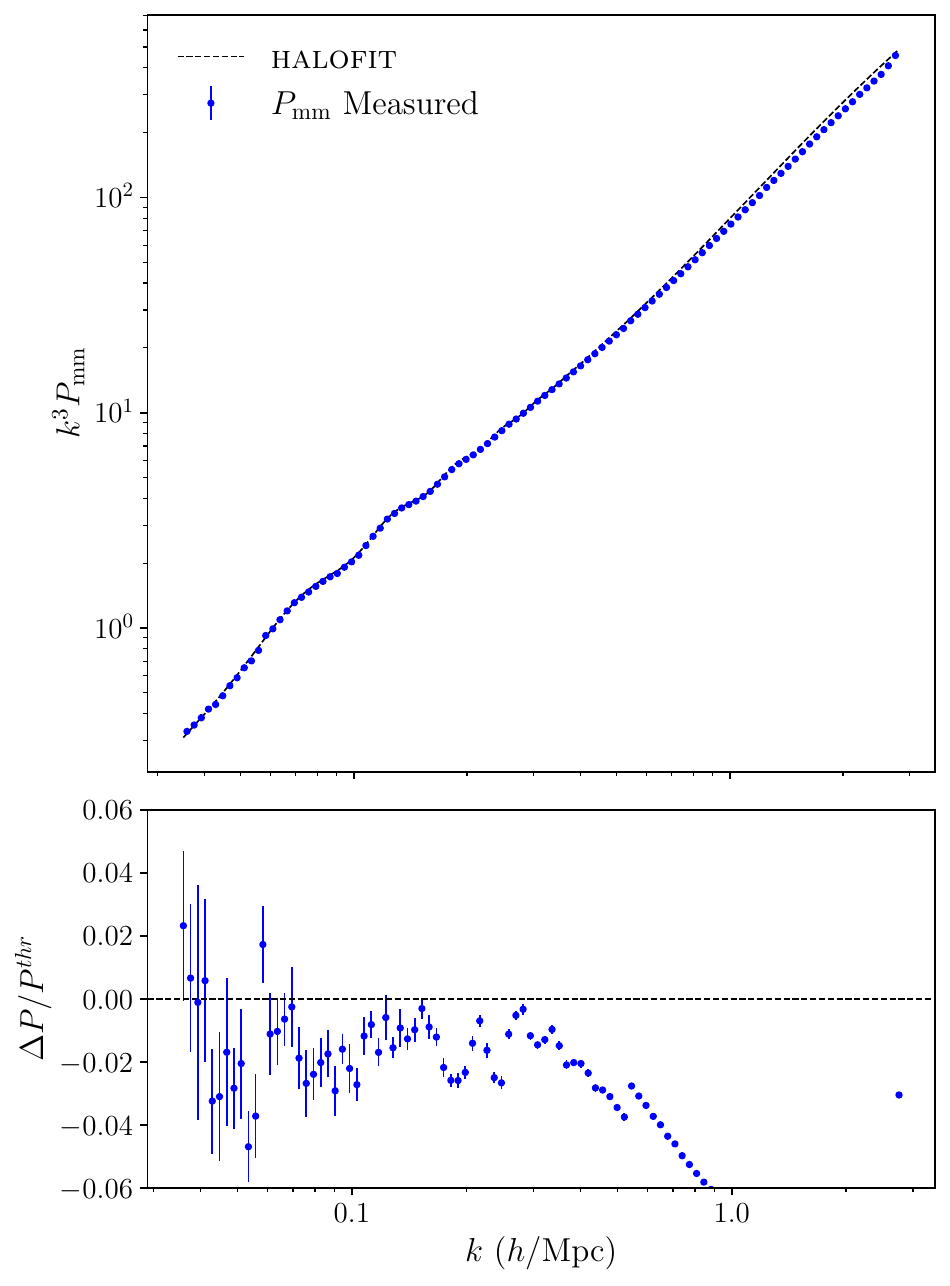}
\caption{Matter power spectrum residuals from the \textsc{halofit} estimate.}
\label{fig:halo_fit_resid}
\end{figure}

In summary, inflating the covariance matrix by 5\% does not have a substantial impact on the main conclusions of this paper. Although the reduced $\chi^2$, which we use almost exclusively as a relative measure of model goodness of fit, changes considerably after inflating the diagonal of the covariance matrix, the percent level accuracy of our model fits and our bias parameter estimates are unaffected. Alongside quantifying the impact of inflating the diagonal of our covariance matrix, we also show that our matter power spectrum errors are consistent with the Gaussian approximation at large scales. We leave the exercise of studying the dependence of the covariance matrix on the number of subboxes to a future study.

\section{Comparison of matter power spectrum with HALOFIT}  \label{Appendix:Halo_Fit}

In Figure \ref{fig:halo_fit_resid} we show the residuals of the computed matter power spectrum and the \textsc{halofit} estimate. We find that \textsc{halofit} and our measurement disagree by approximately 5\% with \textsc{halofit} underestimating our measured matter power spectrum at all scales. These deviations are likely a result of systematics associated with our power spectrum measurement and the underlying accuracy of \textsc{halofit}. We note that inconsistencies between our measured power spectrum and \textsc{halofit} will have a small impact on the conclusions of this analysis in which we considered the ratios of two-point statistics, hence removing any leading order dependence of our results on the \textsc{halofit} estimate.

\section{Poissonian shot-noise results} \label{Appendix:poisson_shotnoise}

\begin{figure*}
\centering
\includegraphics[width=\linewidth]{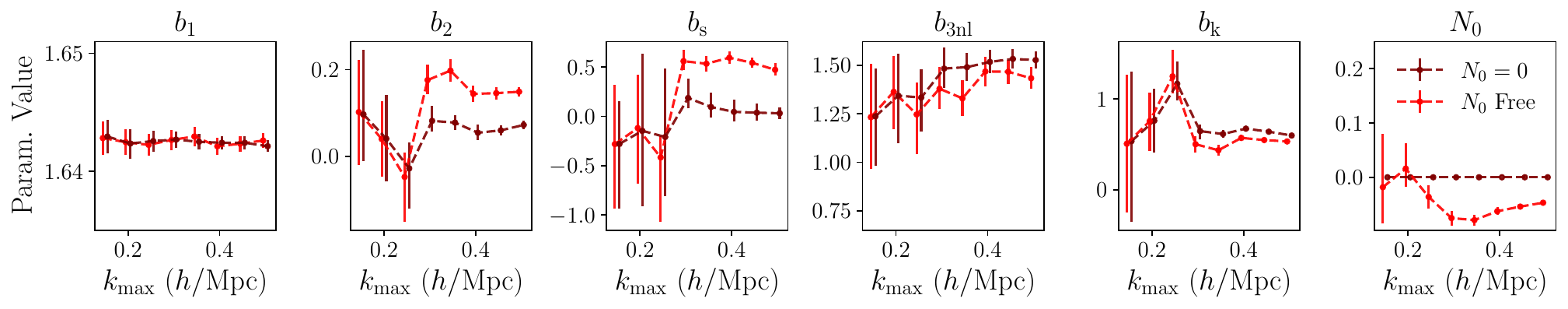}
\caption{Bias parameters as a function of $k_{\rm max}$ for the Fourier space fiducial model fits to the $m_r<23$ galaxy sample with and without including non-Poissonian shot-noise. }
\label{fig:param_fix_SN}
\end{figure*}

 \begin{figure}[!t]
\centering
\includegraphics[width=\linewidth]{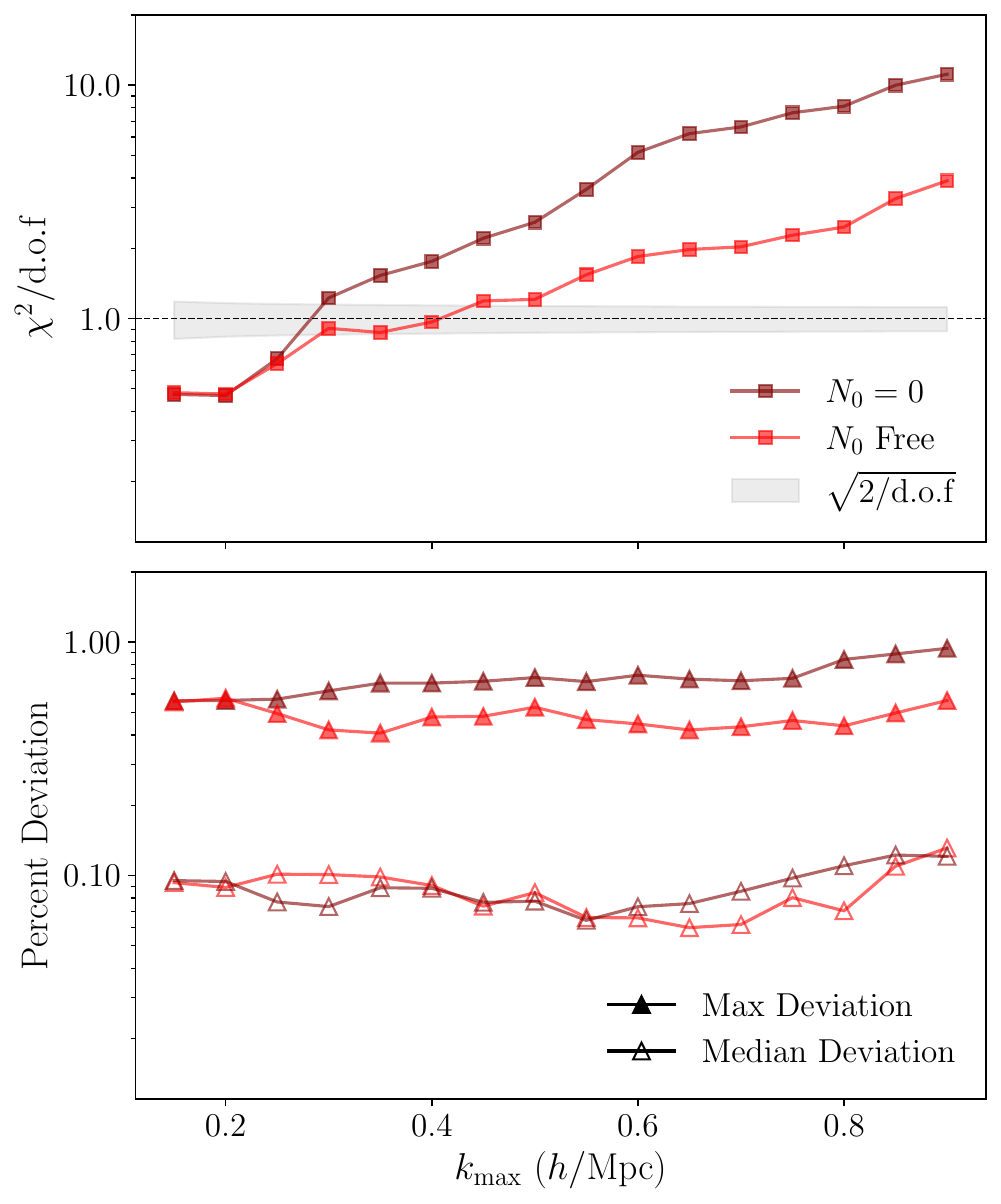}
\caption{\textit{Top:} Reduced $\chi^2$ as a function of $k_{\rm max}$ for the Fourier space fiducial model fits for $m_r<23$ galaxies with and without including non-Poissonian shot-noise. \textit{Bottom:} Maximum and median percent deviation for these model fits.}
\label{fig:chi2_fix_SN}
\end{figure}

In order to assess the impact of including a constant, non-Poissonian shot-noise term in our model, we run a series of fits in which we assume Poissonian shot-noise ($N_0=0$). In Figure \ref{fig:param_fix_SN} we compare bias parameters as a function of $k_{\rm max}$ for the fiducial model fits with Poissonian and non-Poissonian shot-noise to the $m_r<23$ galaxy sample. We find consistent values of $b_1$ across all scale cuts between the two approaches. On the other hand, the non-linear bias parameters can differ significantly between the two models when assuming scale cuts for which the non-Poissonian shot-noise contribution becomes significant ($k_{\rm max}\gtrapprox 0.25$ $h$/Mpc). We note that these changes in the non-linear bias parameters do not resolve the tension between our bias parameter estimates in configuration and Fourier space, hence our result of recovering inconsistent non-linear bias parameters between the two approaches is likely not a consequence of our shot-noise model. We note that assuming Poissonian shot-noise can have a significant impact on the performance of the two parameter model at aggressive scale cuts for which any non-Poissonian contribution will be absorbed by $b_1$ and $b_k$, thus potentially resulting in biased estimates of $b_1$. 

In Figure \ref{fig:chi2_fix_SN} we plot the reduced $\chi^2$, maximum, and median percent deviation as a function of $k_{\rm max}$ for the fiducial model fits to the $m_r<23$ galaxy sample with and without including the non-Poissonian shot-noise term. We find that the reduced $\chi^2$ is noticeably worse when assuming Poissonian shot-noise for $k_{\rm max} \geq 0.3 \ h/{\rm Mpc}$. Nevertheless, the median and maximum percent deviation are consistent between the two models.

\section{Regularization of the configuration space PT model}\label{Appendix:regularization}
\begin{figure*}
\centering
\includegraphics[width=\linewidth]{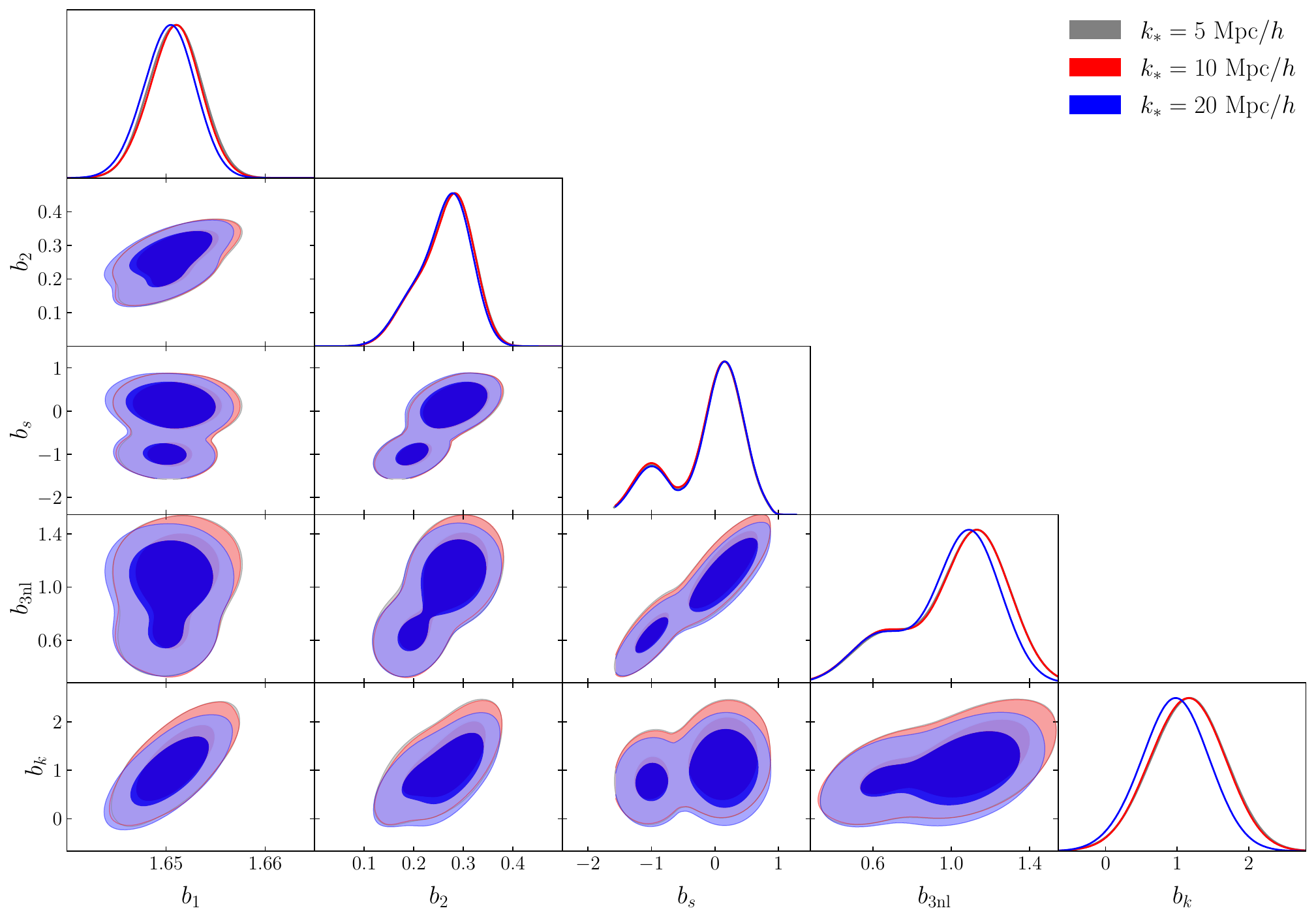}
\caption{Bias parameters posterior distribution when fitting the fiducial model to the ratios $\xi_{\rm gg}/\xi_{\rm mm}$ and $\xi_{\rm gm}/\xi_{\rm mm}$ for the $m_r<23$ galaxy sample with three different values of the regularization scale, $k_*$, assuming $r_{\rm min}=6$ Mpc/$h$.}
\label{fig:regularization}
\end{figure*}

In order to mitigate high-frequency oscillations appearing at large scales in the configuration space model predictions after integrating PT kernels and also regularize the divergent inverse Fourier transform arising from the the higher-derivative bias term we multiply our models for $P_{\rm gg}$ and $P_{\rm gm}$ by an exponential cutoff term, $\exp[-(k/k_*)^4]$, before applying the inverse Fourier transform. In Figure \ref{fig:regularization} we show the bias parameter estimates for the configuration space fiducial model fits at $r_{\rm min}=6$ Mpc/$h$ for three values of $k_*.$ We find  consistent agreement between all bias parameters and conclude that our results are insensitive to the choice of $k_*.$

\begin{figure*}
\centering
\includegraphics[width=\linewidth]{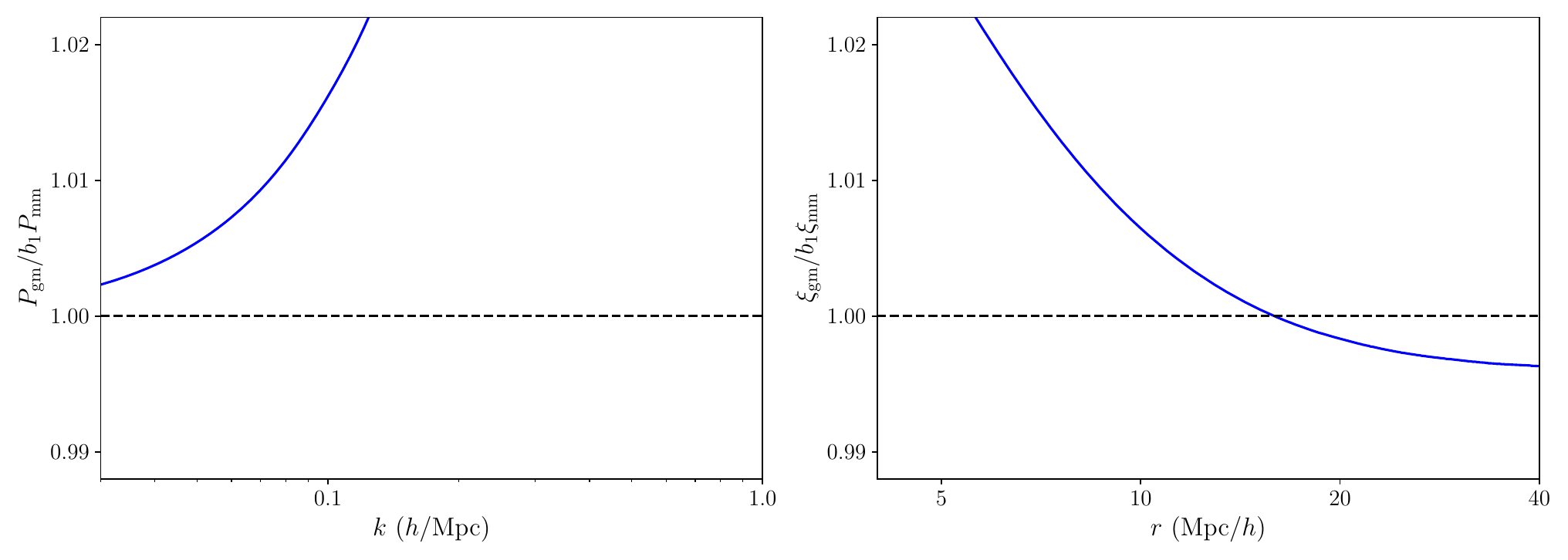}
\caption{Contribution of non-linear terms of the fiducial model to the Fourier space ratios $P_{\rm gm}/b_1P_{\rm mm}$ (left) and configuration space ratios $\xi_{\rm gm}/b_1\xi_{\rm mm}$ (right) generated using the same bias parameters. Linear theory is indicated using a black dashed line.}
\label{fig:linear_bias_comp}
\end{figure*}

\section{Large scale non-linear contributions} \label{Appendix:linear_bias}

In \S\ref{sec:Fourier_config_comp} we found that the linear bias parameter is consistently higher when fitting the correlation functions as opposed to the power spectra. In this section we show that this discrepancy is a result of non-linear terms having sub-percent level contributions to the Fourier space and configuration space data vectors at the largest scales.

Figure \ref{fig:linear_bias_comp} shows the theory predictions for $P_{\rm gm}/b_1P_{\rm mm}$ and $\xi_{\rm gm}/b_1\xi_{\rm mm}$ using the fiducial model and the best fit bias parameter for the $m_r<23$ galaxy sample in Fourier space with $k_{\rm max}=0.35 \ h/{\rm Mpc}.$ The black dashed line indicates the linear contribution. At $k= 0.03 \ h/{\rm Mpc}$, the largest scales we probe in Fourier space, the fiducial model is greater than the linear bias model. On the other hand, at $r= 40 {\rm Mpc}/h$, the largest scales analyzed in configuration space, the fiducial model is less than the linear bias model. In order to compensate for these differences in the non-linear PT contributions, the Fourier space fits will favor lower values of $b_1$ than those in configuration space.

\bibliography{references}

\end{document}